\documentclass[journal]{IEEEtran}

\usepackage{amsmath,amsfonts}
\usepackage[caption=false,font=footnotesize]{subfig}
\usepackage{textcomp}
\usepackage{stfloats}
\usepackage{url}
\usepackage{verbatim}
\usepackage{graphicx}
\usepackage{cite}
\usepackage{setspace}
\usepackage{epsfig}  
\usepackage{amssymb}  
\usepackage{lscape}  
\usepackage{color}
\usepackage{mathrsfs}
\usepackage{mathtools}
\usepackage{amsthm}
\usepackage{array}
\usepackage{booktabs}
 
\usepackage{longtable}
 
\usepackage{cases}
\usepackage{bm}
\usepackage{textcomp}
\usepackage{upgreek}

\usepackage{array}

\begin{document}
%
\title{Multiphysics Numerical Method for Modeling Josephson Traveling-Wave Parametric Amplifiers}
%
%
%

\author{Samuel T. Elkin,~\IEEEmembership{Student Member,~IEEE,}
        Michael Haider,~\IEEEmembership{Member,~IEEE,}
        and~Thomas E. Roth,~\IEEEmembership{Member,~IEEE}
\thanks{Manuscript received XX XX, 2024; revised XX XX, 2024.}
\thanks{Samuel T. Elkin and Thomas E. Roth are with the Elmore Family School of Electrical and Computer Engineering, Purdue University, West Lafayette, IN 47907 USA and the Purdue Quantum Science and Engineering Institute, West Lafayette, IN 47907 USA (e-mail: rothte@purdue.edu)}
\thanks{Michael Haider is with the TUM School of Computation, Information and Technology, Technical University of Munich, 85748 Garching, Germany}}

%
%

\markboth{IEEE Journal on Multiscale and Multiphysics Computational Techniques}%
{Shell \MakeLowercase{\textit{et al.}}: Bare Demo of IEEEtran.cls for IEEE Journals}
%



\maketitle

\begin{abstract}
Josephson traveling-wave parametric amplifiers (JTWPAs) are wideband, ultralow-noise amplifiers used to enable the readout of superconducting qubits. While individual JTWPAs have achieved high performance, behavior between devices is inconsistent due to wide manufacturing tolerances. Amplifier designs could be modified to improve resilience towards variations in amplifier components; however, existing device models often rely on analytical techniques that typically fail to incorporate component variations. To begin addressing this issue, a 1D numerical method for modeling JTWPAs is introduced in this work. The method treats the Josephson junctions and transmission lines in an amplifier as coupled subsystems and can easily incorporate arbitrary parameter variations. We discretize the transmission line subsystem with a finite element time domain method and the Josephson junction subsystem with a finite difference method, with leapfrog time marching used to evolve the system in time. We validate our method by comparing the computed gain to an analytical model for a traditional JTWPA architecture and one with resonant phase matching. We then use our method to demonstrate the impact of variations in Josephson junctions and phase-matching resonators on amplification. In future work, the method will be adjusted to incorporate additional amplifier architectures and extended to a 3D full-wave approach.
\end{abstract}


%
\IEEEpeerreviewmaketitle

\section{Introduction}
\label{sec:intro}
\IEEEPARstart{S}{uperconducting} qubits are a leading architecture in the race towards large-scale quantum computing \cite{2019_Arute_Quantum_Supremacy,wu2021strong,jurcevic2021demonstration}. However, performance improvements are still needed in many of the system components, including those used to measure the states of qubits. To measure a qubit state, an extremely low-power microwave pulse is used to measure the scattering properties of an electromagnetic resonator coupled to the qubit \cite{roth2023transmon,krantz2019quantum}. To perform high-fidelity measurements of the weak microwave signals that carry information about the qubit state, an exceptionally low-noise amplifier (i.e., close to the standard quantum limit) with high gain is required \cite{2017_Walter_Dispersive_Readout, 2020_Aumentado}.

Originally investigated in the late 1960's \cite{1967_Zimmer_ParAmps, 1969_Russer_Paramp}, interest in Josephson parametric amplifiers has been renewed due to their potential for meeting these needs of superconducting qubit technologies \cite{2007_Castellanos_JPA}. These devices use a superconducting Josephson junction to facilitate a nonlinear mixing operation with a strong pump, producing parametric amplification of the desired signal. To improve the achievable gain, the signal is coupled into the nonlinear element through a resonator for high interaction time; however, this comes at the cost of a constrained amplifier bandwidth \cite{2020_Aumentado}. Josephson traveling-wave parametric amplifiers (JTWPAs) were proposed as a wideband alternative, but their lackluster gain combined with challenging manufacturing requirements made them difficult to implement \cite{2021_Esposito_TWPAs}. While these devices can achieve long interaction times by embedding thousands of Josephson junctions in a transmission line, amplification is inhibited by phase mismatches produced through pump-induced dispersion \cite{2014_O'Brien}. However, modern fabrication techniques have enabled the development of resonant phase matching (RPM), where resonant loads are periodically embedded into the transmission line to compensate for deleterious phase mismatch and obtain significantly higher gain \cite{2014_O'Brien, 2015_Macklin}. Over time, the RPM technique has led JTWPAs to become a promising technology for enabling qubit readout due to their wide bandwidth, high gain, and ultra low-noise operation \cite{2021_Esposito_TWPAs, 2015_White_Minimal_RPM, 2017_Grimsmo_JTWPA_RPM, 2023_Qiu_Dual_Pump}. However, alternative techniques for achieving phase matching have also been developed, such as periodically varying the transmission line impedance \cite{2020_Planat_Photonic_JTWPA} and modifying the nonlinearity through flux biasing of Josephson junction-based circuit elements \cite{2022_Ranadive}.

While modern fabrication technologies and design techniques have allowed high performance JTWPAs to be realized, their broader application is still limited by manufacturing variations. Fluctuations between junction properties across a device can produce impedance mismatches, leading to inconsistent performance between JTWPAs \cite{2021_Osman_JJ_reproducability, 2023_Pishchimova, 2020_Bylander_JJ_Viewpoint}. Additionally, even small variations in the resonant loads of RPM architectures can disrupt the delicate phase matching relationship and significantly reduce gain \cite{2020_Feng}. Ideally, JTWPA designers could readily simulate the impact of these variations during the design process and make adjustments to reduce their severity. However, most JTWPA models treat the device as a continuum \cite{2013_Yaakobi_JTWPA_Theory, 2015_Zorin_3-wave, 2020_Planat_Photonic_JTWPA, 2023_Yuan} to develop analytical solutions, preventing variations between elements of the circuit from being modeled.

To combat this issue, previous approaches have accounted for parameter variations by introducing a position-dependent wavenumber \cite{2020_Feng, 2015_Bell_SQUID_JTWPA, 2014_O'Brien}. In these methods, analytical expressions are used to describe segments of a JTWPA with a fixed wavenumber, which are stitched together numerically to simulate the whole JTWPA. While these approaches can accommodate phase mismatch due to parameter variations, they do not incorporate impedance mismatches between segments, which cause reflections that alter performance. As a result, these methods cannot fully capture the impact that manufacturing variability can have on a JTWPA design. Aspects of these effects can be captured in numerical methods using lumped-element circuit simulators such as WRSPICE \cite{2020_Dixon_JTWPA_Complex_Behavior, 2023_Peatain_fab_tolerance, 2023_Guarcello_Numerical}. However, these approaches, along with the others, are not naturally extensible to a 3D full-wave method, preventing them from modeling explicit device geometries. Instead, these methods must estimate circuit parameters from the explicit geometry, which can often be a user-intensive and potentially inaccurate procedure that is avoided in a full-wave approach.

In this work, we develop a 1D multiphysics numerical method which treats the Josephson junctions and transmission lines in a JTWPA as coupled dynamical systems. The transmission lines are discretized via a finite element time domain (FETD) approach, while the differential equations describing junction dynamics are discretized using finite differencing. Interaction between the dynamical variables is facilitated through a leap-frog time-marching procedure. In this treatment, individual JTWPA elements can easily be varied to model arbitrary distributions for the properties of those components. Additionally, this method inherently considers higher-order nonlinear effects such as the generation and impact of harmonics. As JTWPA designs evolve, understanding these effects is increasingly important, but they are difficult to characterize using analytical approaches \cite{2020_Dixon_JTWPA_Complex_Behavior}. Similarly, pump depletion is naturally accounted for in this formulation, allowing the dynamic range of a design to easily be determined. The method can also be extended to a 3D full-wave description in the future to provide superior accuracy in addition to the aforementioned benefits.

Preliminary results on this formulation were reported in \cite{2024_Elkin_AP-S}. This work significantly expands on \cite{2024_Elkin_AP-S} by providing a more detailed formulation of the dynamical equations, as well as extending the method to incorporate RPM. Further, complete details on the numerical discretization approach are also presented. Finally, new numerical results are presented, including a case-study analyzing the impact of process variations in resonant loads and Josephson junctions.

This work is structured as follows. In Section \ref{sec:formulation}, the equations of motion and boundary conditions for a unit cell of a JTWPA are derived, both with and without RPM. Following that, Section \ref{sec:discretization} develops the 1D hybrid method to discretize a unit cell, and extends the description to a full JTWPA for each case. Section \ref{sec:results} validates the method using analytical results for each type of architecture, and demonstrates the effects of variations in the properties of Josephson junctions and resonant loads on amplification. In Section \ref{sec:conclusion}, we discuss the conclusions drawn from these results and the direction for future development of the method.

\section{Formulation}
\label{sec:formulation}
\newcommand{\Lagr}{\mathcal{L}}

\begin{figure}[t!]
\begin{center}
\noindent
  \includegraphics[width=0.9\linewidth]{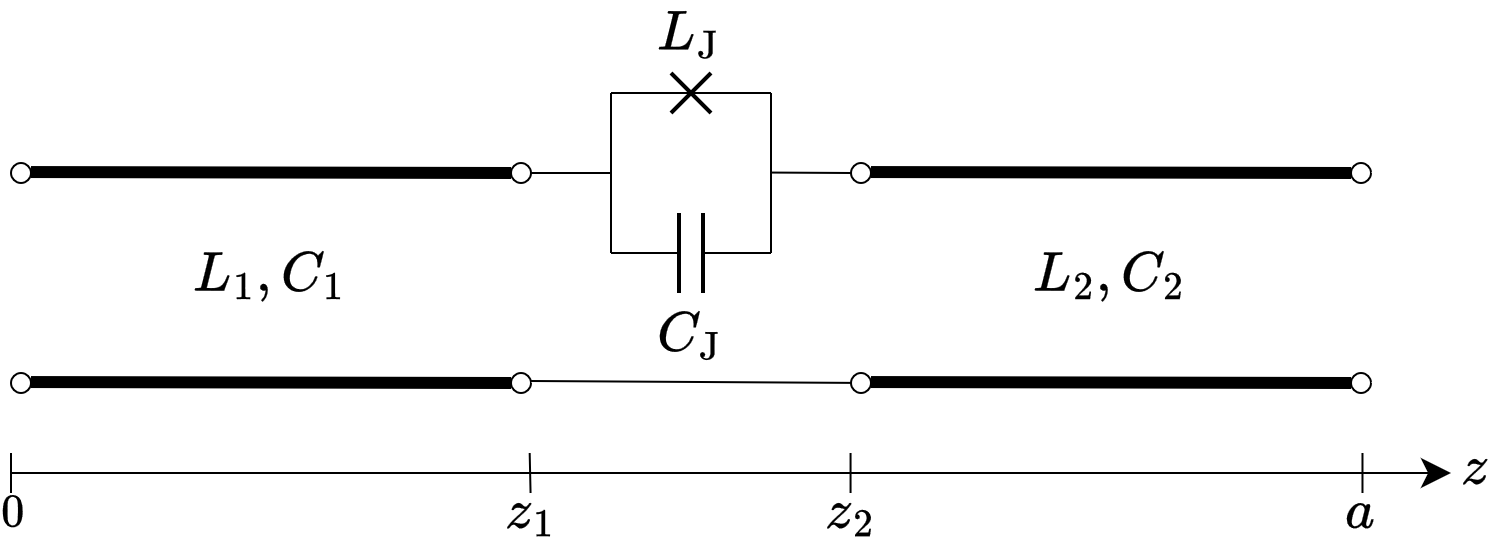}
  \caption{Schematic of a basic single JTWPA unit cell, consisting of a Josephson junction with a nonlinear inductance ($L_{\mathrm{J}}$, denoted by an ``X'') in parallel with a linear capacitance ($C_{\mathrm{J}}$).}
  \label{fig:bc-schematic}
\end{center}
\end{figure}

In this section, the equations of motion (EoMs) and boundary conditions applicable to a single JTWPA unit cell are derived using techniques from the Lagrangian mechanics of continuous systems \cite{stone2009mathematics}. In Section \ref{subsec:basic-JTWPA-cell}, this is done for the ``basic JTWPA unit cell" of Fig. \ref{fig:bc-schematic}, while in Section \ref{subsec:RPM} the derivation is extended to include a resonant load so architectures using RPM can be analyzed.

\subsection{Basic JTWPA Unit Cell}
\label{subsec:basic-JTWPA-cell}

Deriving EoMs and boundary conditions for the unit cell of Fig. \ref{fig:bc-schematic} using Lagrangian mechanics requires a Lagrangian for each portion of the system. Starting with the transmission lines, the Lagrangian for each line can be written as
\begin{align}
    \Lagr_{\mathrm{TX}_i} = \frac{1}{2} \int \big( C_i V^2(z,t) - L_i {I}^2(z,t) \big) \, \mathrm{d}z, 
    \label{eq:tx-lagr-voltage}
\end{align}
where $C_i$ and $L_i$ are the per-unit-length capacitance and inductance of the transmission line. However, to derive EoMs the Lagrangian must be expressed in terms of a generalized coordinate and its derivatives rather than voltage and current. To accomodate this requirement, we use the node flux $\phi(z,t)$, which is related to the voltage and current by $\dot \phi(z,t) = V(z,t)$ and $\phi'(z,t) = -LI(z,t)$, where $\dot \phi(z,t) \equiv \frac{\partial}{\partial t} \phi(z,t)$ and $\phi'(z,t) \equiv \frac{\partial}{\partial z} \phi(z,t)$ \cite{2024_Roth_Semiclassical}. The resulting Lagrangian is
\begin{align}
    \Lagr_{\mathrm{TX}_i} = \frac{1}{2} \int \big( C_i \dot \phi^2(z,t) - L_i^{-1} {\phi'}^2(z,t) \big) \, \mathrm{d}z.
    \label{eq:tx-lagr}
\end{align}

The Josephson junction Lagrangian is developed in \cite{2014_Girvin_Circuit_QED} as
\begin{align}
    \Lagr_{\mathrm{JJ}} = \frac{C_{\mathrm{J}}}{2} \dot{\phi}^2_{\mathrm{J}}(t) + \frac{\hbar I_{\mathrm{c}}}{2e} \cos\bigg(\frac{2e}{\hbar} \phi_{\mathrm{J}}(t)\bigg),
    \label{eq:JJ-lagr}
\end{align}
where $e$ is the elementary charge, $\hbar$ is the reduced Planck's constant, and $\phi_{\mathrm{J}}(t) \equiv \phi(z_1,t) - \phi(z_2,t)$ is the junction flux. The Josephson junction is characterized by the junction capacitance $C_{\mathrm{J}}$ and critical current $I_{\mathrm{c}}$, which is related to the Josephson inductance $L_{\mathrm{J}}$ through $I_{\mathrm{c}} = {\hbar}/(2e L_{\mathrm{J}})$.

To derive the EoMs and boundary conditions in Lagrangian mechanics, we must first introduce the action functional, $S = \int \Lagr \, \mathrm{d}t$. The principle of least action states that a physical system will always operate such that the action is minimized, meaning $\delta S = 0$ \cite{stone2009mathematics}. Typically, the principle of least action can be satisfied using the Euler-Lagrange equation. However, for our continuous systems with non-trivial interconnections as in Fig. \ref{fig:bc-schematic}, the typical Euler-Lagrange equation is no longer applicable. Hence, here we instead take the variation of the action functional directly. We simplify this process by decomposing $\delta S$ into terms for each subsystem,
\begin{align}
    \delta S = \delta S_{\mathrm{TX}_1} + \delta S_{\mathrm{TX}_2} + \delta S_{\mathrm{JJ}}.
    \label{eq:action-decomp}
\end{align}

For the first transmission line, $\delta S_{\mathrm{TX}_1}$ is given by
\begin{multline}
    \delta S_{\mathrm{TX}_1} = \int \bigg( \delta \dot{\phi}(z,t) \frac{\partial \Lagr_{\mathrm{TX}_1}}{\partial \dot{\phi}(z,t)} + \delta \phi'(z,t) \frac{\partial \Lagr_{\mathrm{TX}_1}}{\partial \phi'(z,t)} \bigg) \, \mathrm{d}t.
    \label{eq:var-action-tx1-1}
\end{multline}
By evaluating each derivative using (\ref{eq:tx-lagr}), (\ref{eq:var-action-tx1-1}) becomes 
\begin{multline}
    \delta S_{\mathrm{TX}_1} = \int \int_{0}^{z_1} \big( C_1 \dot \phi(z,t) \delta \dot \phi(z,t) \\ - L_1^{-1} \phi'(z,t) \delta \phi'(z,t) \big) \, \mathrm{d}z \, \mathrm{d}t.
    \label{eq:var-action-tx1-2}
\end{multline}
To remove derivatives from the variations, we employ integration by parts. For the temporal variation $\delta \dot \phi(z,t)$, it is appropriate to consider ``fixed endpoint'' variations so that the boundary terms from the integration by parts can be neglected \cite{stone2009mathematics}. However, for the spatial variation $\delta \phi'(z,t)$, we need to consider variations with ``variable endpoints'' to eventually derive a spatial boundary condition at $z=z_1$ that accounts for the effect of the Josephson junction \cite{stone2009mathematics}. As a result, (\ref{eq:var-action-tx1-2}) becomes
\begin{multline}
    \delta S_{\mathrm{TX}_1} = \int \bigg ( \int_{0}^{z_1}  \big( -C_1 \ddot \phi(z,t) + L_1^{-1} \phi''(z,t) \big) \delta \phi(z,t) \, \mathrm{d}z \\ - L_1^{-1} \phi'(z_1,t) \delta \phi(z_1,t) \bigg) \, \mathrm{d}t,
    \label{eq:var-action-tx1}
\end{multline}
Similarly, for the other transmission line we find $\delta S_{\mathrm{TX}_2}$ is
\begin{multline}
    \delta S_{\mathrm{TX}_2} = \int \bigg ( \int_{z_2}^{a}  \big( -C_2 \ddot \phi(z,t) + L_2^{-1} \phi''(z,t) \big) \delta \phi(z,t) \, \mathrm{d}z \\ + L_2^{-1} \phi'(z_2,t) \delta \phi(z_2,t) \bigg) \, \mathrm{d}t.
    \label{eq:var-action-tx2}
\end{multline}

\begin{figure}[t!]
\begin{center}
\noindent
  \includegraphics[width=0.9\linewidth]{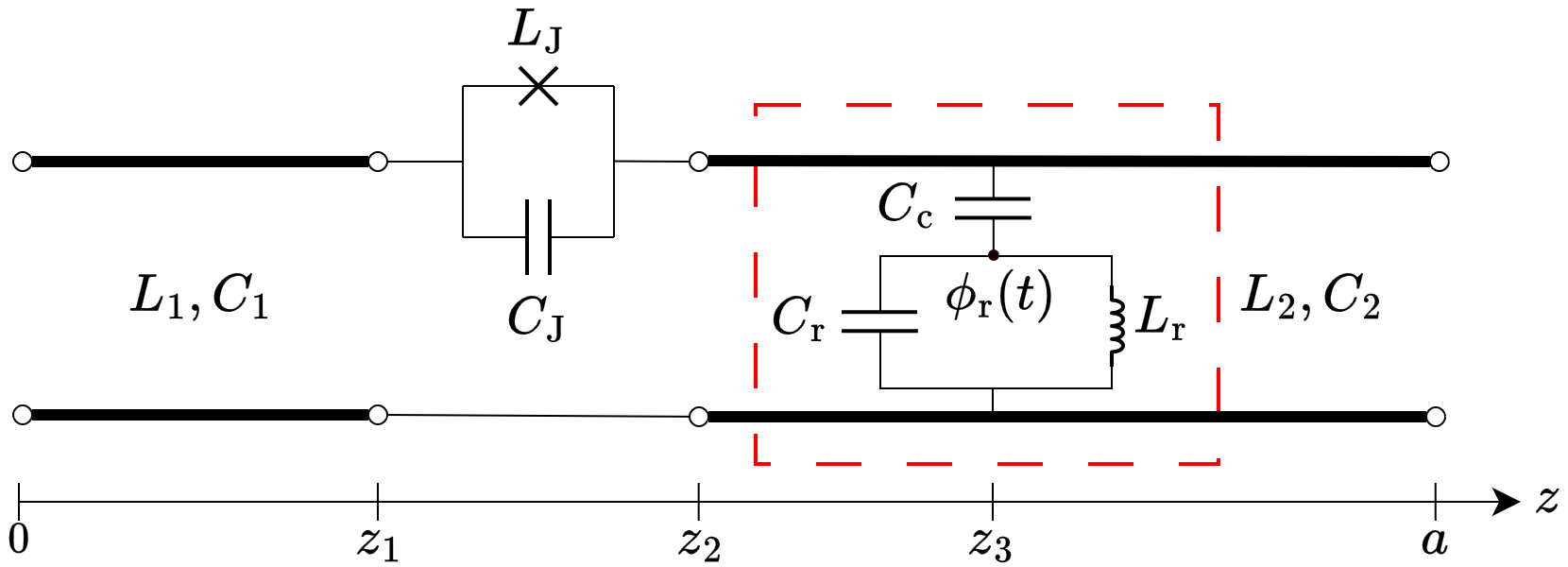}
  \caption{Schematic of a JTWPA unit cell for a RPM design. The added resonant load and coupling capacitor are outlined in red.}
  \label{fig:bc-schematic-RPM}
\end{center}
\end{figure}

Moving to the Josephson junction, the change of independent variables in (\ref{eq:JJ-lagr}) leads to the expression
\begin{align}
    \delta S_{\mathrm{JJ}} = \int \bigg( \delta \phi_{\mathrm{J}}(t) \frac{\partial \Lagr}{\partial \phi_{\mathrm{J}}(t)} + \delta \dot{\phi}_{\mathrm{J}}(t) \frac{\partial \Lagr}{\partial \dot{\phi}_{\mathrm{J}}(t)} \bigg) \, \mathrm{d}t.
\end{align}
By evaluating the derivatives using (\ref{eq:JJ-lagr}), we obtain
\begin{multline}
    \delta S_{\mathrm{JJ}} = \int \bigg[ C_{\mathrm{J}} \dot \phi_{\mathrm{J}}(t) \big(\delta \dot \phi (z_1,t) - \delta \dot \phi (z_2,t) \big) \\ - I_{\mathrm{c}} \sin \bigg( \frac{2e}{\hbar} \phi_{\mathrm{J}}(t) \bigg) \big( \delta \phi(z_1,t) - \delta \phi(z_2,t) \big) \bigg] \, \mathrm{d}t,
    \label{eq:var-action-JJ-2}
\end{multline}
where substitutions were made for $\delta \phi_{\mathrm{J}}(t)$ and $\delta \dot \phi_{\mathrm{J}}(t)$ using $\phi_{\mathrm{J}}(t) = \phi(z_1,t) - \phi(z_2,t)$. Integrating by parts produces
\begin{multline}
    \delta S_{\mathrm{JJ}} = -\int \bigg[ C_{\mathrm{J}} \ddot \phi_{\mathrm{J}}(t) \big(\delta \phi (z_1,t) - \delta \phi (z_2,t) \big) \\ + I_{\mathrm{c}} \sin \bigg( \frac{2e}{\hbar} \phi_{\mathrm{J}}(t) \bigg) \big( \delta \phi(z_1,t) - \delta \phi(z_2,t) \big) \bigg] \, \mathrm{d}t,
    \label{eq:var-action-JJ}
\end{multline}
which is used alongside (\ref{eq:var-action-tx1}) and (\ref{eq:var-action-tx2}) to evaluate (\ref{eq:action-decomp}),
\begin{multline}
    \delta S = \int \bigg[ \int_{0}^{z_1}  \big( -C_1 \ddot \phi(z,t) + L_1^{-1} \phi''(z,t) \big) \delta \phi(z,t) \, \mathrm{d}z \\ - \bigg( L_1^{-1} \phi'(z_1,t) + C_{\mathrm{J}} \ddot \phi_{\mathrm{J}}(t) + I_{\mathrm{c}} \sin \big(\frac{2e}{\hbar} \phi_{\mathrm{J}}(t) \big) \bigg) \delta \phi(z_1, t) \\ + \bigg( L_2^{-1} \phi'(z_2,t) + C_{\mathrm{J}} \ddot \phi_{\mathrm{J}}(t) + I_{\mathrm{c}} \sin \big(\frac{2e}{\hbar} \phi_{\mathrm{J}}(t) \big) \bigg) \delta \phi(z_2, t) \\ + \int_{z_2}^{a}  \big( -C_2 \ddot \phi(z,t) + L_2^{-1} \phi''(z,t) \big) \delta \phi(z,t) \, \mathrm{d}z \bigg] \mathrm{d}t.
    \label{eq:var-action}
\end{multline}
The principle of least action states that $\delta S = 0$ for all $t$. Since each variation is nonzero by definition, the coefficient of each variation in (\ref{eq:var-action}) must equate to zero. This step produces the equations of motion and boundary conditions,
\begin{align}
    \phi''(z,t) - L_1 C_1 \ddot \phi(z,t) = 0, \quad 0<z<z_1, 
    \label{eq:EoM-tx1} \\ 
    L_1^{-1} \phi'(z_1,t) = -C_{\mathrm{J}} \ddot \phi_{\mathrm{J}}(t) - I_{\mathrm{c}} \sin \bigg(\frac{2e}{\hbar} \phi_{\mathrm{J}}(t) \bigg), 
    \label{eq:EoM-JJ1} \\
    L_2^{-1} \phi'(z_2,t) = - C_{\mathrm{J}} \ddot \phi_{\mathrm{J}}(t) - I_{\mathrm{c}} \sin \bigg(\frac{2e}{\hbar} \phi_{\mathrm{J}}(t) \bigg), \label{eq:EoM-JJ2} \\
    \phi''(z,t) - L_2 C_2 \ddot \phi(z,t) = 0, \quad z_2<z<a, 
    \label{eq:EoM-tx2}
\end{align}


Clearly, (\ref{eq:EoM-tx1}) and (\ref{eq:EoM-tx2}) describe waves propagating down each transmission line. The boundary conditions (\ref{eq:EoM-JJ1}) and (\ref{eq:EoM-JJ2}) introduce nonlinearity through interactions with the Josephson junction. To move towards allowing this nonlinear set of equations to be solved with a linear time-marching procedure, we treat $\phi_{\mathrm{J}}(t)$ as a second dynamical variable with EoM
\begin{multline}
    C_{\mathrm{J}} \ddot \phi_{\mathrm{J}}(t) + I_{\mathrm{c}} \sin \bigg(\frac{2e}{\hbar} \phi_{\mathrm{J}}(t) \bigg) \\ = -\frac{1}{2L_1} \phi'(z_1,t) - \frac{1}{2L_2} \phi'(z_2,t).
    \label{eq:phi-J-time-march}
\end{multline}
This is obtained by summing (\ref{eq:EoM-JJ1}) and (\ref{eq:EoM-JJ2}). A numerical procedure to enable a linear solution for time-marching the dynamical variables in (\ref{eq:EoM-tx1}), (\ref{eq:EoM-tx2}), and (\ref{eq:phi-J-time-march}) together will be discussed in Section \ref{sec:discretization}.

\subsection{RPM JTWPA Unit Cell}
\label{subsec:RPM}

To implement RPM, resonant loads of capacitance $C_{\mathrm{r}}$ and inductance $L_{\mathrm{r}}$ are periodically coupled into unit cells through capacitance $C_{\mathrm{c}}$, as shown in Fig. \ref{fig:bc-schematic-RPM}. To determine how this change impacts the equations of motion, we reconstruct the Lagrangian for the second transmission line, which becomes 
\begin{multline}
    \Lagr_{\mathrm{TX}_2} = \frac{1}{2} \int_{z_2}^{a} \big( C_2 \dot \phi^2(z,t) - L_2^{-1} {\phi'}^2(z,t) \big) \, \mathrm{d}z \\ + \frac{1}{2} C_{\mathrm{c}} \big( \dot \phi(z_3,t) - \dot \phi_{\mathrm{r}}(t) \big)^2\! + \frac{1}{2} \big( C_{\mathrm{r}} \dot \phi_{\mathrm{r}}^2(t) - L_{\mathrm{r}}^{-1}  {\phi}^2_{\mathrm{r}}(t)\big),
    \label{eq:RPM-lagr}
\end{multline}
where $\phi_{\mathrm{r}}(t)$ is the flux stored in the resonator. Therefore, the variation of the action functional is given by
\begin{multline}
    \delta S_{\mathrm{TX}_2} = \int \bigg( \delta \phi_{\mathrm{r}}(z,t) \frac{\partial \Lagr}{\partial \phi_{\mathrm{r}}(z,t)} \\ + \delta \dot{\phi_{\mathrm{r}}}(z,t) \frac{\partial \Lagr}{\partial \dot{\phi_{\mathrm{r}}}(z,t)} + \delta \phi'(z,t) \frac{\partial \Lagr}{\partial \phi'(z,t)} \\ + \delta \dot{\phi}(z,t) \frac{\partial \Lagr}{\partial \dot{\phi}(z,t)} + \delta \dot{\phi}(z_3,t) \frac{\partial \Lagr}{\partial \dot{\phi}(z_3,t)} \bigg) \, \mathrm{d}t.
    \label{eq:var-action-rpm-tx2-1}
\end{multline}
Substituting (\ref{eq:RPM-lagr}) into (\ref{eq:var-action-rpm-tx2-1}) produces the expression
\begin{multline}
    \delta S_{\mathrm{TX}_2} = \int \bigg( \int_{z_2}^{a} \big( C_1 \dot \phi(z,t) \delta \dot \phi(z,t) \\ - L_1^{-1} \phi'(z,t) \delta \phi'(z,t) \big) \, \mathrm{d}z \\ + C_{\mathrm{r}} \dot \phi_{\mathrm{r}}(t) \delta \dot \phi_{\mathrm{r}}(t) - L_{\mathrm{r}}^{-1}\phi_{\mathrm{r}}(t) \delta \phi_{\mathrm{r}}(t) \\ + C_{\mathrm{c}} \big(\dot \phi(z_3,t) - \dot \phi_{\mathrm{r}}(t) \big) \big(\delta \dot \phi(z_3, t) - \delta \dot \phi_{\mathrm{r}}(t) \big) \bigg) \, \mathrm{d}t,
    \label{eq:var-action-rpm-tx2-2}
\end{multline}
which after integrating by parts becomes
\begin{multline}
    \delta S_{\mathrm{TX}_2} = \int \bigg[ \int_{z_2}^{a}  \big( -C_2 \ddot \phi(z,t) + L_2^{-1} \phi''(z,t) \big) \delta \phi(z,t) \, \mathrm{d}z \\ + L_2^{-1} \phi'(z_2,t) \delta \phi(z_2,t) \\ - \bigg( C_{\mathrm{r}} \ddot \phi_{\mathrm{r}}(t) - L_{\mathrm{r}}^{-1} \phi_{\mathrm{r}}(t) + C_{\mathrm{c}} \big( \ddot \phi(z_3,t) - \ddot \phi_{\mathrm{r}}(t) \big) \bigg) \delta \phi_{\mathrm{r}}(t) \\ - C_{\mathrm{c}} \big( \ddot \phi(z_3,t) - \ddot \phi_{\mathrm{r}}(t) \big) \delta \phi(z_3,t) \bigg] \mathrm{d}t.
    \label{eq:var-action-rpm-tx2}
\end{multline}

This expression can be used in conjunction with (\ref{eq:var-action-tx1}) and (\ref{eq:var-action-JJ}) to determine $\delta S$ for the whole RPM unit cell,
\begin{multline}
    \delta S = \int \bigg[ \int_{0}^{z_1}  \big( -C_1 \ddot \phi(z,t) + L_1^{-1} \phi''(z,t) \big) \delta \phi(z,t) \, \mathrm{d}z \\ + \int_{z_2}^{a}  \big( -C_2 \ddot \phi(z,t) + L_2^{-1} \phi''(z,t) \big) \delta \phi(z,t) \, \mathrm{d}z \\ - \bigg( L_1^{-1} \phi'(z_1,t) + \frac{C_{\mathrm{J}}}{2} \ddot \phi_{\mathrm{J}}(t) + I_{\mathrm{c}} \sin \big(\frac{2e}{\hbar} \phi_{\mathrm{J}}(t) \big) \bigg) \delta \phi(z_1, t) \\ + \bigg( L_2^{-1} \phi'(z_2,t) + \frac{C_{\mathrm{J}}}{2} \ddot \phi_{\mathrm{J}}(t) + I_{\mathrm{c}} \sin \big(\frac{2e}{\hbar} \phi_{\mathrm{J}}(t) \big) \bigg) \delta \phi(z_2, t) \\ - \bigg( C_{\mathrm{r}} \ddot \phi_{\mathrm{r}}(t) - L_{\mathrm{r}}^{-1} \phi_{\mathrm{r}}(t) + C_{\mathrm{c}} \big( \ddot \phi(z_3,t) - \ddot \phi_{\mathrm{r}}(t) \big) \bigg) \delta \phi_{\mathrm{r}}(t) \\ - C_{\mathrm{c}} \big( \ddot \phi(z_3,t) - \ddot \phi_{\mathrm{r}}(t) \big) \delta \phi(z_3,t) \bigg] \mathrm{d}t.
    \label{eq:var-action-RPM}
\end{multline}
Applying the principle of least action to (\ref{eq:var-action-RPM}) leads to
\begin{gather}
    \phi''(z,t) - L_1 C_1 \ddot \phi(z,t) = 0, \quad 0<z<z_1, 
    \label{eq:EoM-tx1-RPM} \\ 
    L_1^{-1} \phi'(z_1,t) = -C_{\mathrm{J}} \ddot \phi_{\mathrm{J}}(t) - I_{\mathrm{c}} \sin \bigg(\frac{2e}{\hbar} \phi_{\mathrm{J}}(t) \bigg), 
    \label{eq:EoM-JJ1-RPM} \\
    L_2^{-1} \phi'(z_2,t) = - C_{\mathrm{J}} \ddot \phi_{\mathrm{J}}(t) - I_{\mathrm{c}} \sin \bigg(\frac{2e}{\hbar} \phi_{\mathrm{J}}(t) \bigg), 
    \label{eq:EoM-JJ2-RPM} \\
    \phi''(z,t) - L_2 C_2 \ddot \phi(z,t) = 0, \quad z_2<z<z_3,
    \label{eq:EoM-tx2-RPM-1} \\ 
    \phi''(z_3,t) - L_2 (C_2+C_{\mathrm{c}}) \ddot \phi(z_3,t) = -L_2 C_{\mathrm{c}} \ddot \phi_{\mathrm{r}}(t),
    \label{eq:EoM-tx2-RPM-loc} \\
    \phi''(z,t) - L_2 C_2 \ddot \phi(z,t) = 0, \quad z_3<z<a,
    \label{eq:EoM-tx2-RPM-2} \\ 
    \phi_{\mathrm{r}}(t) - L_{\mathrm{r}} (C_{\mathrm{r}} + C_{\mathrm{c}}) \ddot \phi_{\mathrm{r}}(t) = -L_{\mathrm{r}} C_{\mathrm{c}} \ddot \phi(z_3,t).
    \label{eq:RPM-time-march}
\end{gather}
Since (\ref{eq:EoM-JJ1-RPM}) and (\ref{eq:EoM-JJ2-RPM}) are unchanged from (\ref{eq:EoM-JJ1}) and (\ref{eq:EoM-JJ2}), (\ref{eq:phi-J-time-march}) still holds for the RPM unit cell. A numerical procedure to enable a linear solution for time-marching the dynamical variables in (\ref{eq:phi-J-time-march}), (\ref{eq:EoM-tx1-RPM}), and (\ref{eq:EoM-tx2-RPM-1}) to (\ref{eq:RPM-time-march})  together will be discussed in Section \ref{sec:discretization}.

\section{Discretization}
\label{sec:discretization}
Here, we detail the development of the hybrid model from the EoMs for each unit cell. In this method, 1D FETD is applied to discretize $\phi(z,t)$ along the transmission lines. FETD is chosen to simplify the transition to a 3D full-wave implementation in the future. Temporal discretization of $\phi_{\mathrm{J}}(t)$ is performed using central differencing, which will be demonstrated to produce a linear time-marching equation. The direct relationship between the dynamical variables is eliminated using leap-frog time-marching, which allows $\phi(z,t)$ to be evaluated without the need for a nonlinear solver.

This section is structured as follows. First, the method for the basic JTWPA unit cell is developed in Section \ref{subsec:basic-cell-disc} by deriving time-marching equations for the dynamical variables. Additionally, we introduce the leap-frog time-stepping procedure to linearly advance the dynamical variables in time. In Section \ref{subsec:full-JTWPA-sim}, the unit cells are combined and integrated with driving and terminating components to model a full JTWPA. Finally, the time-marching equations are adjusted to accomodate the addition of resonant loads to model a RPM JTWPA in Section \ref{subsec:rpm-cell-disc}.

\subsection{Basic JTWPA Unit Cell}
\label{subsec:basic-cell-disc}

To begin, we discretize the node flux $\phi(z,t)$ along the transmission lines, which is described by (\ref{eq:EoM-tx1}) to (\ref{eq:EoM-tx2}). We follow a standard FETD process outlined in \cite{2010_Jin_CEM,2024_Roth_Semiclassical} with first-order (triangular) nodal basis functions $N_i(z)$ such that
\begin{align}
    \phi(z,t) = \sum_i \phi_i(t) N_i(z).
    \label{eq:phi-disc}
\end{align}
This process leads to a matrix equation of the form
\begin{align}
   [T_{\mathrm{u}}] \frac{d^2}{dt^2} \{ \phi(t) \} + [S_{\mathrm{u}}] \{ \phi(t) \} = \{ f_{\mathrm{u}}(t) \},
   \label{eq:line-diff-eq}
\end{align}
where $\{ \phi(t) \}_i = \phi_i(t)$ and
\begin{multline}
        \relax [T_{\mathrm{u}}]_{ij} = L_1 C_1 \int_0^{z_1} N_i(z) N_j(z) \, \mathrm{d}z \\ + L_1 C_{\mathrm{J}} \big( \delta_{jM} - \delta_{j(M+1)} \big) \delta_{iM} + L_2 C_2 \int_{z_2}^{a} N_i(z) N_j(z) \, \mathrm{d}z  \\ - L_2 C_{\mathrm{J}} \big( \delta_{jM} - \delta_{j(M+1)} \big) \delta_{i(M+1)}, 
        \label{eq:T-basic-FETD}
\end{multline}
\begin{align}
    [S_{\mathrm{u}}]_{ij} = \int_0^{z_1} N_i'(z) N_j'(z) \, \mathrm{d}z + \int_{z_2}^{a} N_i'(z) N_j'(z) \, \mathrm{d}z , 
    \label{eq:S-basic-FETD} \\
    \{ f_{\mathrm{u}}(t) \}_{i} = I_{\mathrm{c}} \big( L_2 \delta_{i(M+1)} - L_1 \delta_{iM} \big) \sin \bigg( \frac{2e}{\hbar} \phi_{\mathrm{J}}(t) \bigg),
    \label{eq:f-basic-FETD}
\end{align}
where $\delta_{ij}$ is a Kronecker delta, and nodes $M$ and $M+1$ are positioned at $z_1$ and $z_2$ respectively. For the junction flux, applying (\ref{eq:phi-disc}) leads (\ref{eq:phi-J-time-march}) to become
\begin{multline}
    C_{\mathrm{J}} \ddot \phi_{\mathrm{J}}(t) + I_{\mathrm{c}} \sin \bigg(\frac{2e}{\hbar} \phi_{\mathrm{J}}(t) \bigg) \\ = \sum_i \bigg( -\frac{1}{2L_1} N_i'(z_1) - \frac{1}{2L_2} N_i'(z_2) \bigg) \phi_i(t) \\ = -\frac{1}{2L_1 \Delta z} \big(\phi_M(t) - \phi_{M-1}(t) \big) \\ - \frac{1}{2L_2 \Delta z} \big( \phi_{M+2}(t) - \phi_{M+1}(t) \big),
    \label{eq:phi-J-time-march-disc}
\end{multline}
where the final equality was reached by taking piecewise derivatives of the triangular basis functions.

At this point, $\{ \phi(t) \}$ and $\phi_{\mathrm{J}}(t)$ are still directly related through $\phi_{\mathrm{J}}(t) \equiv \phi_{M}(t) - \phi_{M+1}(t)$, which invalidates the source treatment of $\phi_{\mathrm{J}}(t)$ in (\ref{eq:f-basic-FETD}). However, leap-frogging between the dynamical variables in the time discretization can be exploited to circumvent this issue. More explicitly, by solving for each dynamical variable at alternating half-timesteps, the direct relationship between them is no longer enforced, enabling the ``source treatment'' for $\phi_{\mathrm{J}}(t)$ in (\ref{eq:f-basic-FETD}). Applying this method using central differencing yields time-marching equations for each dynamical variable,
\begin{multline}
    \relax [T_{\mathrm{u}}] \{ \phi^{(n+1)} \} = \frac{(\Delta t)^2}{2} \big(\{ f_{\mathrm{u}}^{(n+1/2)} \} + \{ f_{\mathrm{u}}^{(n-1/2)} \} \big) \\ - \big( (\Delta t)^2 [S_{\mathrm{u}}] - 2 [T_{\mathrm{u}}] \big) \{ \phi^{(n)} \} - [T_{\mathrm{u}}] \{ \phi^{(n-1)} \}
    \label{eq:time-march-phi}
\end{multline}
and
\begin{multline}
    \phi_{\mathrm{J}}^{(n+3/2)} = 2 \phi_{\mathrm{J}}^{(n+1/2)} - \frac{I_{\mathrm{c}} (\Delta t)^2}{C_{\mathrm{J}}} \sin \bigg( \frac{2e}{\hbar} \phi_{\mathrm{J}}^{(n+1/2)} \bigg) \\ - \phi_{\mathrm{J}}^{(n-1/2)} - \frac{(\Delta t)^2}{4 L_1 C_{\mathrm{J}} \Delta z} \big( \phi_{M}^{(n+1)} + \phi_{M}^{(n)} - \phi_{M-1}^{(n+1)} - \phi_{M-1}^{(n)} \big) \\ - \frac{(\Delta t)^2}{4 L_2 C_{\mathrm{J}} \Delta z} \big( \phi_{M+2}^{(n+1)} + \phi_{M+2}^{n} - \phi_{M+1}^{(n+1)} - \phi_{M+1}^{n} \big),
    \label{eq:time-march-phiJ}
\end{multline}
where $\phi_M^{(n)}$ is the flux at node $M$ and timestep $n$. In formulating (\ref{eq:time-march-phi}) and (\ref{eq:time-march-phiJ}), averaging was used to approximate unknown values, i.e.: $\phi_M^{(n+1/2)} \approx \frac{1}{2} \big( \phi_M^{(n+1)} + \phi_M^{(n)} \big)$. Notably, using central differencing to discretize (\ref{eq:phi-J-time-march-disc}) led to a time-marching equation for $\phi_{\mathrm{J}}(t)$ that can be solved linearly, despite the nonlinearity of (\ref{eq:phi-J-time-march-disc}). As a result, both dynamical variables can be evolved in time without the need for a nonlinear solver.

\subsection{Full JTWPA Simulation}
\label{subsec:full-JTWPA-sim}

\begin{figure}[t!]
\begin{center}
\noindent
  \includegraphics[width=1.0\linewidth]{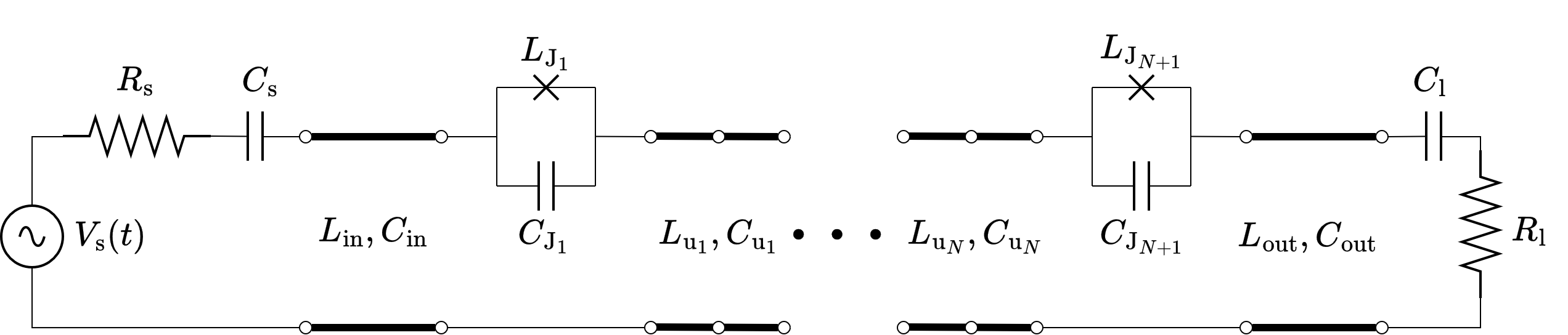}
  \caption{Schematic of a JTWPA. Capacitive loads are introduced at the source and termination of the JTWPA to aid in filtering out low-frequency spurious modes that lead to late-time instability.}
  \label{fig:JTWPA}
\end{center}
\end{figure}

The method is extended from a single unit cell to a JTWPA by connecting $N$ unit cells with per-unit-length inductance $L_{\mathrm{u}_n}$ and capacitance $C_{\mathrm{u}_n}$ together at their endpoints, as shown in Fig. \ref{fig:JTWPA}. To drive the JTWPA, a voltage source $V_{\mathrm{s}}(t)$ with resistance $R_{\mathrm{s}}$ is connected to the the input side through a transmission line with per-unit-length inductance $L_{\mathrm{in}}$ and capacitance $C_{\mathrm{in}}$. On the output side, a load $R_{\mathrm{l}}$ is connected through another transmission line characterized by per-unit-length parameters $L_{\mathrm{out}}$ and $C_{\mathrm{out}}$ to terminate the simulation region. The source and load capacitors $C_{\mathrm{s}}$ and $C_{\mathrm{l}}$ are introduced to aid in filtering out low-frequency spurious modes that lead to late-time instability, which will be discussed in more detail in Section \ref{sec:results}. The boundary conditions for the source and load are introduced in \cite{2024_Roth_Semiclassical} as
\begin{align}
    \phi'(z_{\mathrm{s}},t) = -\frac{L_{\mathrm{in}}}{R_{\mathrm{s}}} [ V_{\mathrm{s}}(t) - \dot \phi(z_{\mathrm{s}},t)], \\
    \phi'(z_{\mathrm{l}},t) = -\frac{L_{\mathrm{out}}}{R_{\mathrm{l}}} \dot \phi(z_{\mathrm{l}},t), 
\end{align}
where $z_{\mathrm{s}}$ and $z_{\mathrm{l}}$ are the positions of the source and load. Since these terms contain a first derivative in time, the matrix equation of (\ref{eq:line-diff-eq}) needs to be modified to the form
\begin{align}
   [T] \frac{d^2}{dt^2} \{ \phi(t) \} + [R] \frac{d}{dt} \{ \phi(t) \} + [S] \{ \phi(t) \} = \{ f(t) \}.
   \label{eq:line-diff-eq-full}
\end{align}
The matrices for the full JTWPA in this expression become
\begin{align}
    \relax [T] = [T_{\mathrm{in}}] + \sum_n [T_{\mathrm{u}_n}] + [T_{\mathrm{out}}], 
    \label{eq:T-full} \\
    [R]_{ij} = \frac{L_{\mathrm{in}}}{R_{\mathrm{s}}} \delta_{i1} \delta_{j1} + \frac{L_{\mathrm{out}}}{R_{l}} \delta_{i N_e} \delta_{i N_e}, 
    \label{eq:R-full} \\
    [S] = [S_{\mathrm{in}}] + \sum_n [S_{\mathrm{u}_n}] + [S_{\mathrm{out}}], 
    \label{eq:S-full} \\
    \{f(t)\}_i = \frac{L_{\mathrm{in}}}{R_{\mathrm{s}}} V_{\mathrm{s}}(t) \delta_{i1} + \sum_n \{f_{\mathrm{u}_n}(t)\},
    \label{eq:f-full}
\end{align}
where node $1$ occurs at the source, and node $N_e$ is positioned at the load. Further, $[T_{\mathrm{u}_n}]$, $[S_{\mathrm{u}_n}]$, and $\{f_{\mathrm{u}_n}(t)\}$ refer to (\ref{eq:T-basic-FETD}), (\ref{eq:S-basic-FETD}), and (\ref{eq:f-basic-FETD}) for unit cell $n$. The input matrices have the form
\begin{multline}
    [T_{\mathrm{in}}]_{ij} = L_{\mathrm{in}} C_{\mathrm{in}} \int_{z_{\mathrm{s}}}^{z_0} N_i(z) N_j(z) \, \mathrm{d}z \\ + L_{\mathrm{in}} C_{\mathrm{s}} \big( \delta_{11} - \delta_{12} - \delta_{21} + \delta_{22} \big),
\end{multline}
\begin{align}
    [S_{\mathrm{in}}]_{ij} = \int_{z_{\mathrm{s}}}^{z_0} N_i'(z) N_j'(z) \, \mathrm{d}z,
\end{align}
where $z_0$ is the starting position of the first unit cell. The output matrices can be expressed as
\begin{multline}
    [T_{\mathrm{out}}]_{ij} = L_{\mathrm{out}} C_{\mathrm{out}} \int_{z_{\mathrm{e}}}^{z_{\mathrm{l}}} N_i(z) N_j(z) \, \mathrm{d}z \\ + L_{\mathrm{out}} C_{\mathrm{l}} \big( \delta_{(N_e-1) (N_e-1)} - \delta_{(N_e-1)N_e} \\ - \delta_{N_e(N_e-1)} + \delta_{N_e N_e} \big),
\end{multline}
\begin{align}
    [S_{\mathrm{out}}]_{ij} = \int_{z_{\mathrm{e}}}^{z_{\mathrm{l}}} N_i'(z) N_j'(z) \, \mathrm{d}z,
\end{align}
where $z_{\mathrm{e}}$ is the end position of the final unit cell.

With the matrices in (\ref{eq:line-diff-eq-full}) constructed, central differencing can be applied to produce a time-marching equation,
\begin{multline}
    \relax \bigg( [T] + \frac{\Delta t}{2} [R] \bigg) \{ \phi^{(n+1)} \} = (\Delta t)^2 \{ f^{(n)} \} \\ + \frac{(\Delta t)^2}{2} \big(\{ f^{(n+1/2)} \} + \{ f^{(n-1/2)} \} \big) \\ - \big( (\Delta t)^2 [S] - 2 [T] \big) \{ \phi^{(n)} \} - \bigg( [T] - \frac{\Delta t}{2} [R] \bigg) \{ \phi^{(n-1)} \}.
    \label{eq:time-march-phi-full}
\end{multline}
The time marching equation for $\phi_{\mathrm{J}}(t)$ in (\ref{eq:time-march-phiJ}) is unchanged for the full JTWPA, but now must be independently solved for each Josephson junction in the circuit.

\subsection{RPM JTWPA}
\label{subsec:rpm-cell-disc}

The inclusion of resonant loads introduces the resonator flux $\phi_{\mathrm{r}}(t)$ as a third dynamical variable for an RPM JTWPA. However, since (\ref{eq:RPM-time-march}) is linear, there is no need to incorporate leap-frogging to evolve $\phi_{\mathrm{r}}(t)$ in time. Instead, the resonator flux $\phi_{\mathrm{r}}(t)$ can be solved simultaneously with $\phi(z,t)$. To accomodate this change, we modify (\ref{eq:line-diff-eq-full}) to the form 
\begin{multline}
    \begin{bmatrix}
        T_{\ell \ell} & T_{\ell \mathrm{r}} \\
        T_{\mathrm{r} \ell} & T_{\mathrm{r} \mathrm{r}}
    \end{bmatrix}
    \frac{d^2}{dt^2}
    \begin{Bmatrix}
        \phi(t) \\
        \phi_{\mathrm{r}}(t)
    \end{Bmatrix}
    +
    \begin{bmatrix}
        R_{\ell \ell} & 0 \\
        0 & 0 \\
    \end{bmatrix}
    \frac{d}{dt}
    \begin{Bmatrix}
        \phi(t) \\
        \phi_{\mathrm{r}}(t)
    \end{Bmatrix} 
    \\ +
    \begin{bmatrix}
        S_{\ell \ell} & 0 \\
        0 & S_{\mathrm{r} \mathrm{r}}
    \end{bmatrix}
    \begin{Bmatrix}
        \phi(t) \\
        \phi_{\mathrm{r}}(t)
    \end{Bmatrix}
    = 
    \begin{Bmatrix} 
        f(t) \\
        0
    \end{Bmatrix},
    \label{eq:line-diff-eq-RPM}
\end{multline}
where $\{ \phi_{\mathrm{r}} (t) \}_i = \phi_{\mathrm{r},i} (t)$, or the flux in the resonant load of the $i$th unit cell. Further, $[R]_{\ell \ell}$, $[S]_{\ell \ell}$, and $\{ f(t) \}$ are equivalent to their counterparts in (\ref{eq:R-full}) to (\ref{eq:f-full}) for the full JTWPA without RPM. From (\ref{eq:EoM-tx1-RPM}) to (\ref{eq:RPM-time-march}), the remaining matrices can be expressed as
\begin{gather}
    [T_{\ell \ell}]_{ij} = [T]_{ij} + \sum_n L_{\mathrm{u}_n} C_{\mathrm{c}_n} \delta_{iR_n} \delta_{jR_n} \\
    [T_{\ell \mathrm{r}}]_{ij} = \sum_n -L_{\mathrm{u}_n} C_{\mathrm{c}_n} \delta_{iR_n} \delta_{jn}, \\
    [T_{\mathrm{r} \ell}]_{ij} = \sum_n -L_{\mathrm{r}_n} C_{\mathrm{c}_n} \delta_{in} \delta_{jR_n}, \\
    [T_{\mathrm{r} \mathrm{r}}]_{ij} = \sum_n L_{\mathrm{r}_n} (C_{\mathrm{c}_n} + C_{\mathrm{r}_n}) \delta_{in} \delta_{jn}, \\
    [S_{\mathrm{r} \mathrm{r}}] = \sum_n \delta_{in} \delta_{jn},
\end{gather}
where each node $R_n$ is where the resonant load is located in the $n$th unit cell, and $L_{\mathrm{r}_n}$, $C_{\mathrm{c}_n}$, and $C_{\mathrm{r}_n}$ are $L_{\mathrm{r}}$, $C_{\mathrm{c}}$, and $C_{\mathrm{r}}$ for that unit cell. 

The time-marching expression for the RPM JTWPA is the same as the equation for the basic JTWPA in (\ref{eq:time-march-phi-full}), but with the matrices and vectors modified to the block form of (\ref{eq:line-diff-eq-RPM}). The time-marching equation for each $\phi_{\mathrm{J}}(t)$ in (\ref{eq:time-march-phiJ}) is unchanged by the addition of resonant loads.

\section{Results}
\label{sec:results}
Here, we present results to validate the numerical method and exhibit its utility. In Section \ref{subsec:gain-curve}, the gain of a basic JTWPA design is calculated from the results of the numerical method and compared to analytical results. This procedure is extended to a JTWPA architecture using RPM in Section \ref{subsec:gain-curve-RPM}, which also includes a discussion of the strategy used to inhibit late-time instability. In Section \ref{subsec:JJ-variations}, the method is applied to simulate the effect of variations in Josephson junction area across a JTWPA. Further, the resonant loads are varied in Section \ref{subsec:RPM-variations} to demonstrate another potential application.

\subsection{Basic JTWPA Gain Curve}
\label{subsec:gain-curve}

\begin{figure}[t!]
\begin{center}
\noindent
  \includegraphics[width=0.9\linewidth]{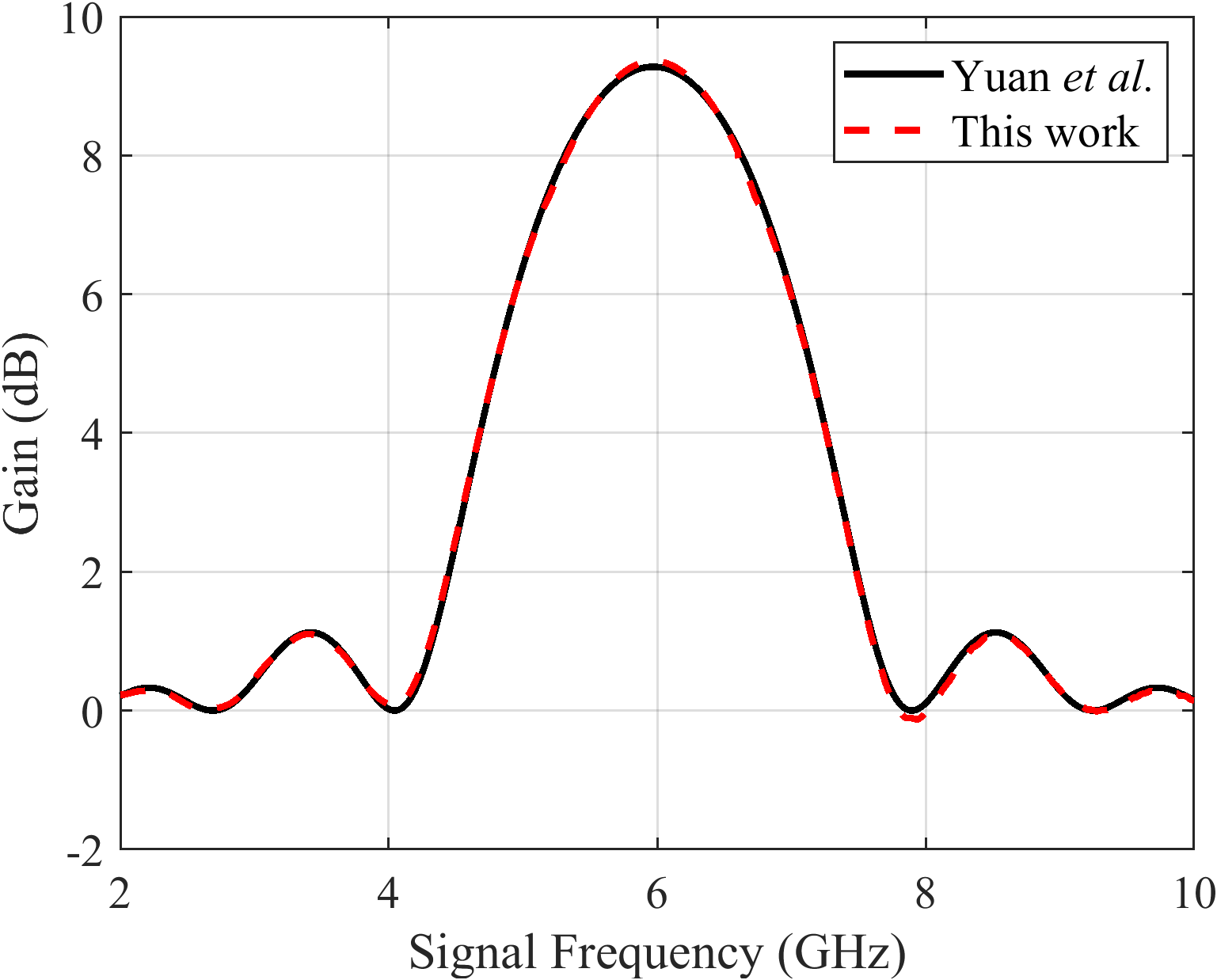}
  \caption{Gain of the architecture of Table \ref{table:JTWPA-params} calculated using the analytical method of \cite{2023_Yuan} and the numerical method of this work. Points between $5.3 \, \mathrm{GHz}$ and $6.6 \, \mathrm{GHz}$ are interpolated using a quadratic fit.}
  \label{fig:gain-curve}
\end{center}
\end{figure}

\begin{table}[b]
\caption{Parameters for the basic JTWPA architecture from \cite{2014_O'Brien}} 
\renewcommand{\arraystretch}{1.5}
\centering
\newcolumntype{D}{>{\centering\arraybackslash} m{0.1\textwidth}}
\newcolumntype{M}{>{\centering\arraybackslash} m{0.325\textwidth}}
\footnotesize
\label{table:JTWPA-params}
\begin{tabular}{M|D}
 \hline
 \hline
 \textbf{Parameter} & \textbf{Value}\\
 \hline
 \hline
 Junction critical current ($I_{\mathrm{c}}$) & $3.29 \, \upmu \mathrm{A}$\\
 \hline
 Josephson capacitance ($C_{\mathrm{J}}$) & $329 \, \mathrm{fF}$\\
 \hline
 Per-unit-length capacitance ($C_{\mathrm{u}}$) & $3.9 \, \mathrm{nF/m}$\\
 \hline
 Unit cell length ($a$) & $10 \, \upmu \mathrm{m}$ \\
 \hline
 Number of unit cells ($N$) & $2000$ \\
 \hline
 Pump current magnitude ($I_{\mathrm{p}}$) & $1.645 \, \upmu \mathrm{A}$ \\
 \hline
 Pump frequency ($f_{\mathrm{p}}$) & $5.970  \, \mathrm{GHz}$ \\
 \hline
 \hline
\end{tabular}
\end{table}

The method is initially validated by comparing to the analytical results from \cite{2023_Yuan} for the basic JTWPA design presented in \cite{2014_O'Brien}. The specifications for this JTWPA are detailed in Table~\ref{table:JTWPA-params}. Notably, the inductance of each unit cell in this design is assumed to be dominated by $L_{\mathrm{J}}$, meaning $L_{\mathrm{u}} \approx 0$. This assumption is not suitable for our numerical method, so a small per-unit-length inductance of $L_{\mathrm{u}} = 1 \, \upmu\mathrm{H/m}$ is added to the transmission lines. This small $L_{\mathrm{u}}$ was found to have a minimal impact on the numerical results, while also not modifying the FETD stability constraint to a problematic extent. To match the input and output transmission lines to the JTWPA, the contributions of the Josephson junctions need to be accounted for, so $L_{\mathrm{in}} = L_{\mathrm{out}} = L_{\mathrm{u}} + L_{\mathrm{J}}/a$.

The JTWPA is driven with a voltage source
\begin{align}
    V_{\mathrm{s}}(t) = 2 I_{\mathrm{p}} Z_0 W(t) \big( \sin( \omega_{\mathrm{p}} t) + \alpha \sin( \omega_{\mathrm{s}} t) \big),
    \label{eq:source-voltage}
\end{align}
where $\omega_{\mathrm{s}}$ is the signal frequency, $\omega_{\mathrm{p}}$ is the pump frequency, and $Z_0 = (aL_{\mathrm{u}} + L_{\mathrm{J}})/aC_{\mathrm{u}}$ is the intrinsic impedance of the transmission line. The ratio between the signal and pump amplitudes is determined by $\alpha$, which is set to $\alpha = 1/20$ to prevent pump depletion from compressing gain. Further, $W(t)$ is an exponential tapering function with a total width of $50 \, \mathrm{ns}$. While analytical models typically assume the pump and signal are monochromatic, due to our method operating in the time domain a finite bandwidth is necessary to keep the total simulation duration at a reasonable level.

The gain $G$ is calculated from simulation results via
\begin{align}
    G = 10 \log \bigg( \frac{\int_{f_{\mathrm{L}}}^{f_{\mathrm{H}}} |\phi_{\mathrm{o}}(f)|^2 \, \mathrm{d}f}{\int_{f_{\mathrm{L}}}^{f_{\mathrm{H}}} |\phi_{\mathrm{i}}(f)|^2 \, \mathrm{d}f}\bigg), 
    \label{eq:gain}
\end{align}
where $\phi_{\mathrm{i}}(f)$ and $\phi_{\mathrm{o}}(f)$ are the Fourier transform of the node flux at the input and output respectively. The integration bandwidth is determined by $f_{\mathrm{L}}$ and $f_{\mathrm{H}}$, which are set to $f_{\mathrm{L}} = f_{\mathrm{s}} - 0.5 \, \mathrm{GHz}$ and $f_{\mathrm{H}} = f_{\mathrm{s}} + 0.5 \, \mathrm{GHz}$. Wideband integration is necessary due to self-mixing of the signal, which spreads energy from its center frequency over a wider frequency range. In an experimental setting, the signal is typically several orders of magnitude smaller than the critical current, which limits this self-mixing process. In the numerical method, this process becomes relevant due to the relatively large signal-to-pump ratio, which was chosen to allow an increased pump bandwidth that helps reduce the total simulation duration to keep the method more efficient.

Using (\ref{eq:gain}), the gain is extracted from numerical simulations for the architecture of Table \ref{table:JTWPA-params} and compared to the analytical result of \cite{2023_Yuan} in Fig. \ref{fig:gain-curve}. At signal frequencies between $5.3 \, \mathrm{GHz}$ and $6.6 \, \mathrm{GHz}$, the integration bandwidth overlaps with the pump and induces error in the gain calculation. To allow the gain to be extracted in this region, a quadratic fit is used to interpolate it from surrounding points. While interpolation is necessary for this topology, alternative JTWPA designs using three-wave mixing remove the pump from the operating bandwidth, allowing this process to be avoided. These architectures will be discussed in more detail in Section \ref{sec:conclusion} and considered in future work.

\subsection{RPM JTWPA Gain Curve}
\label{subsec:gain-curve-RPM}

\begin{table}[b]
\caption{Parameters for the RPM JTWPA architecture from \cite{2014_O'Brien}} 
\label{table:JTWPA-RPM-params}
\renewcommand{\arraystretch}{1.5}
\centering
\newcolumntype{D}{>{\centering\arraybackslash} m{0.1\textwidth}}
\newcolumntype{M}{>{\centering\arraybackslash} m{0.325\textwidth}}
\footnotesize
\begin{tabular}{M|D}
 \hline
 \hline
 \textbf{Parameter} & \textbf{Value}\\
 \hline
 \hline
 Junction critical current ($I_{\mathrm{c}}$) & $3.29 \, \upmu \mathrm{A}$\\
 \hline
 Josephson capacitance ($C_{\mathrm{J}}$) & $329 \, \mathrm{fF}$\\
 \hline
 Per-unit-length capacitance ($C_{\mathrm{u}}$) & $3.9 \, \mathrm{nF/m}$\\
 \hline
 Unit cell length ($a$) & $10 \, \upmu \mathrm{m}$ \\
 \hline
 Resonant load coupling capacitance ($C_{\mathrm{c}}$) & $10 \, \mathrm{fF}$\\
 \hline
 Resonant load capacitance ($C_{\mathrm{r}}$) &  $7.036 \, \mathrm{pF}$\\
 \hline
 Resonant load inductance ($L_{\mathrm{r}}$) & $100 \, \mathrm{pH}$\\
 \hline
 Number of unit cells ($N$) & $2000$ \\
 \hline
 Pump current magnitude ($I_{\mathrm{p}}$) & $1.645 \, \upmu \mathrm{A}$ \\
 \hline
 Pump frequency ($f_{\mathrm{p}}$) & $5.970  \, \mathrm{GHz}$ \\
 \hline
 \hline
\end{tabular}
\end{table}

\begin{figure}[t!]
\centering
  \includegraphics[width=0.9\linewidth]{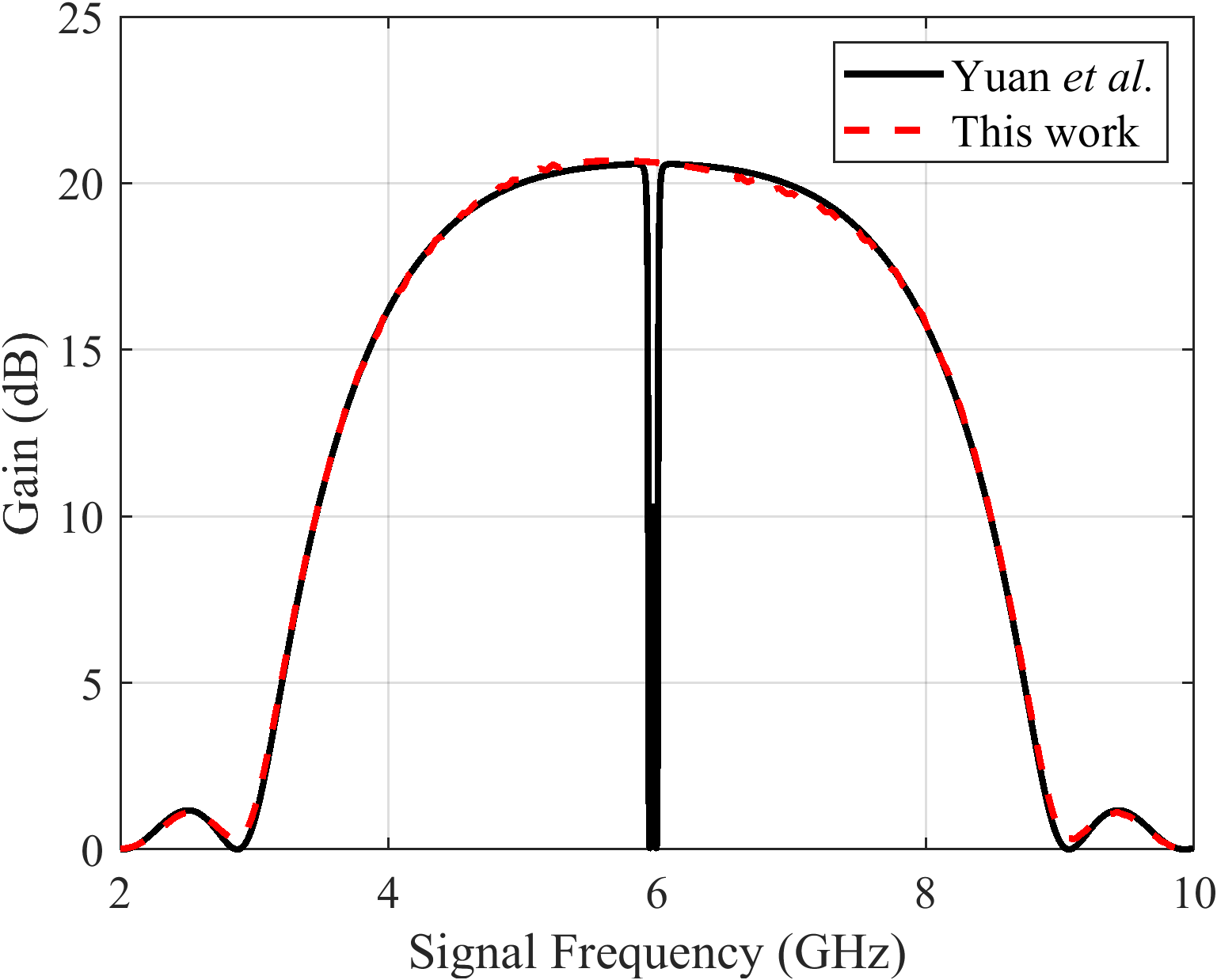}
  \caption{Gain of the architecture of Table \ref{table:JTWPA-RPM-params} calculated using the analytical method of \cite{2023_Yuan} and the numerical method of this work. Points between $5.3 \, \mathrm{GHz}$ and $6.6 \, \mathrm{GHz}$ are interpolated using a fourth-order polynomial fit.}
  \label{fig:gain-curve-RPM}
\end{figure}

The method is also validated using the RPM JTWPA design in \cite{2014_O'Brien}, for which the relevant parameter values are listed in Table \ref{table:JTWPA-RPM-params}. For this architecture, some additional changes to the voltage source described in (\ref{eq:source-voltage}) must be made. First of all, the addition of the resonant loads shifts the average intrinsic impedance of the JTWPA to $Z_0 = (aL_{\mathrm{u}} + L_{\mathrm{J}})/(aC_{\mathrm{u}} + C_{\mathrm{c}})$. To compensate for this, the per-unit-length capacitance of the input and output transmission lines must also be adjusted to $C_{\mathrm{in}} = C_{\mathrm{out}} = C_{\mathrm{u}} + C_{\mathrm{c}}/a$. Additionally, the increased gain of this design requires that the signal-to-pump ratio is reduced to $\alpha = 1/50$ to minimize the effect of pump depletion. Finally, the pump bandwidth must be reduced for this architecture due to the addition of the resonant loads. To correct for mismatch between the signal and pump phases, the resonant loads introduce a pole at $6 \, \mathrm{GHz}$, where its proximity to the pump leads to a phase shift. However, if the pump is too wideband, this proximity means the sidebands of the pump will overlap with the pole, resulting in deformation. In an experimental setting, the pump is typically narrowband enough that this is not an issue. However, the shortened envelope used in the numerical method increases the pump bandwidth, causing it to extend into the pole. To minimize this issue, the width of the tapering function $W(t)$ was increased to $250 \, \mathrm{ns}$. 

While the narrowband pump is necessary in this architecture, increasing the duration of $W(t)$ requires an equivalent increase in total simulation time. As a result, low-frequency spurious solutions are given time to grow and eventually produce instability. To address this issue, we applied the correction method detailed in \cite{2007_Chilton_Late_Time_Stability}. This method periodically isolates and removes the electrostatic portion of the solution to prevent buildup of spurious solutions. To separate this electrostatic component, a basis for the nullspace of $[S]$ must be known. For the 1D system, each vector in the basis is constant over a single segment of transmission line terminated by Josephson junctions or capacitors, and zero elsewhere.

\begin{figure*}[t!]
    \centering
    \subfloat[][]{
    \includegraphics[width=0.47\linewidth]{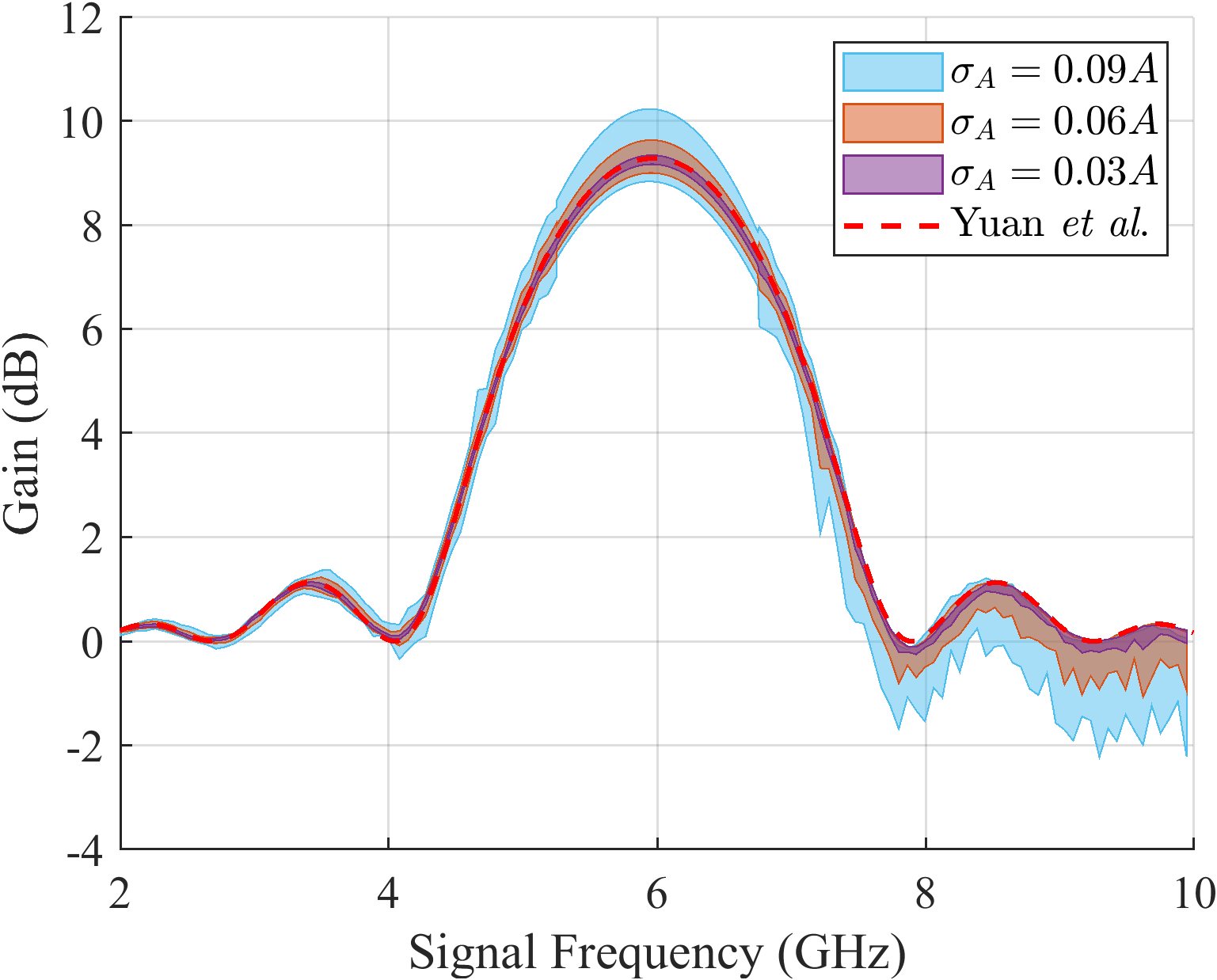}
    \label{subfig:noRPM-areaGrad=0}
    }
    \subfloat[][]{
    \includegraphics[width=0.47\linewidth]{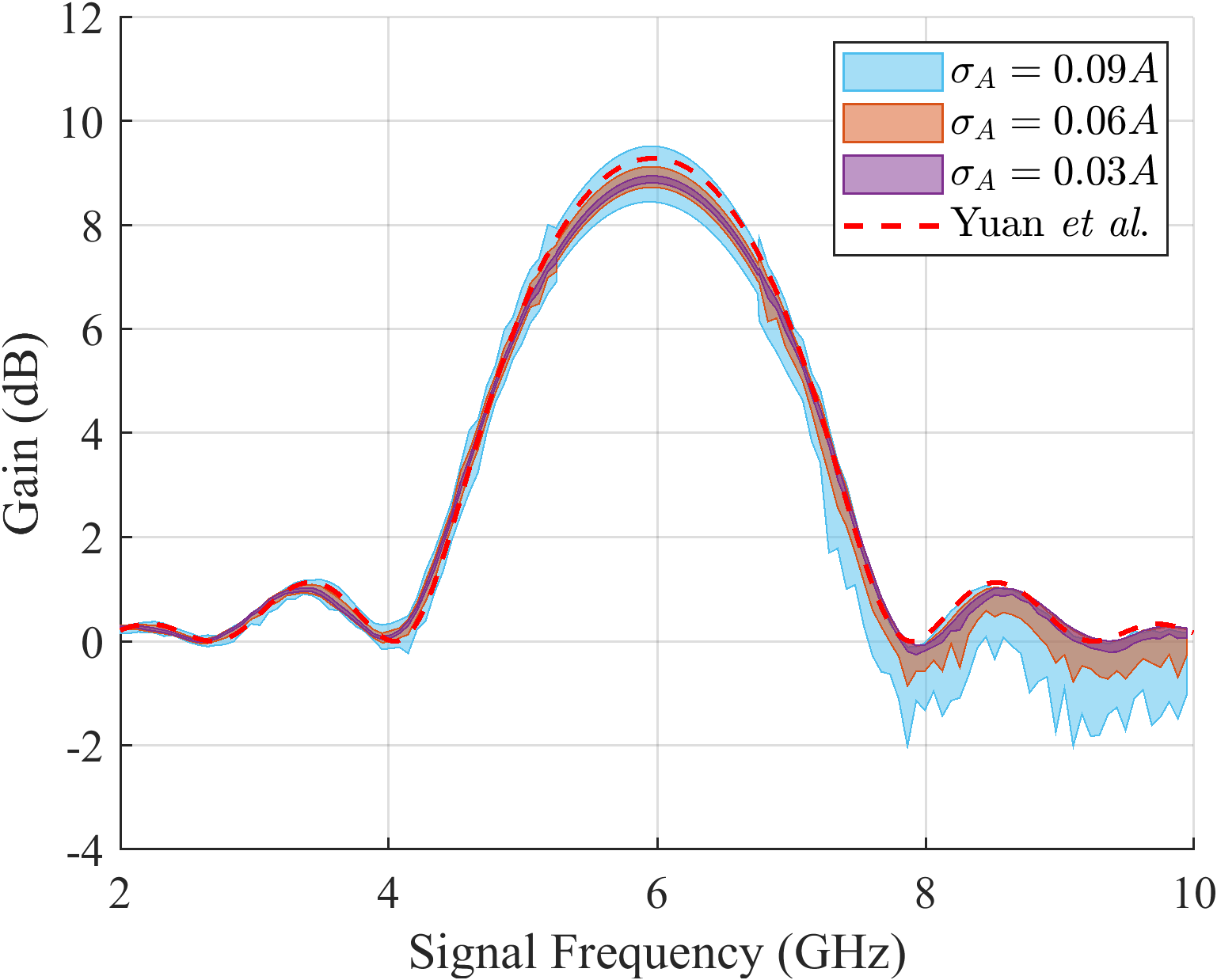}  
    \label{subfig:noRPM-areaGrad=5}
    }
    \hfill
    \subfloat[][]{
    \includegraphics[width=0.47\linewidth]{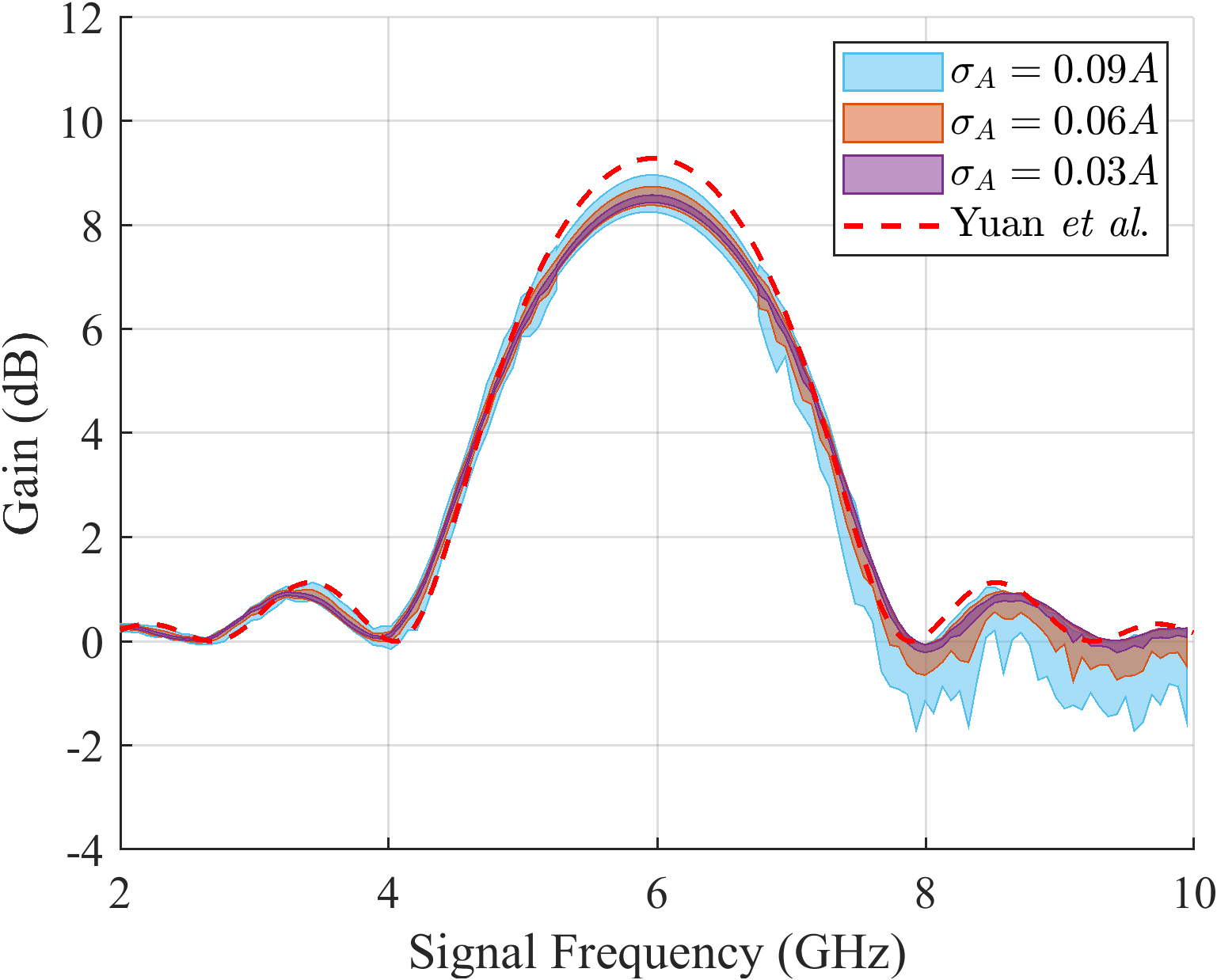}  
    \label{subfig:noRPM-areaGrad=10}
    }
    \subfloat[][]{
    \includegraphics[width=0.47\linewidth]{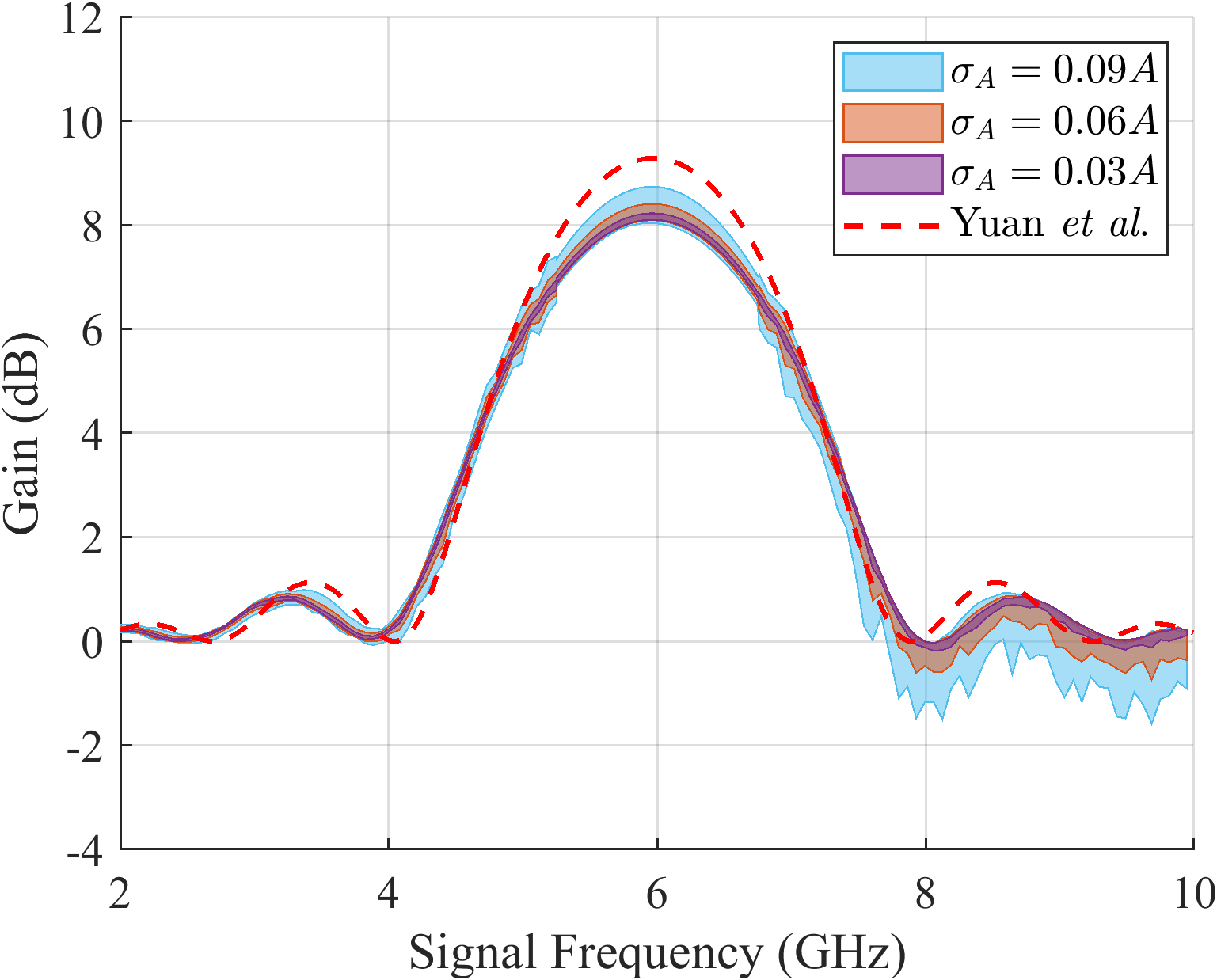}  
    \label{subfig:noRPM-areaGrad=15}
    }
    \caption{Gain of the Table \ref{table:JTWPA-params} JTWPA with variations in Josephson junction area. Each curve is shaded between the minimum and maximum outcomes of 50 simulations. The area of each junction in the JTWPA is independently selected on a normal distribution with standard deviation $\sigma_A$. The mean of the distribution is (a) constant, or linearly varied by (b) $0.05A$, (c) $0.10A$, or (d) $0.15A$ across the JTWPA.}
    \label{fig:JTWPA-noRPM-areaVar}
\end{figure*}

Importantly, the correction method of \cite{2007_Chilton_Late_Time_Stability} is designed to be performed on lossless systems. While the JTWPA itself is lossless, $R_{\mathrm{s}}$ and $R_{\mathrm{l}}$ introduce loss which can produce inaccuracy in the correction process. To avoid needing to apply the method in these regions, capacitors $C_{\mathrm{s}}$ and $C_{\mathrm{l}}$ are introduced. These capacitors separate the source and load from the input and output transmission lines, which removes them from the corresponding nullspace vectors. As a result, the correction method can be applied to the input and output transmission lines without also applying it to the source and load. To prevent instability from building up in the uncorrected regions and spreading to the JTWPA, the capacitors are sized at $C_{\mathrm{s}} = C_{\mathrm{l}} = 100 \, \mathrm{pF}$. At this value, they present a high impedance to spurious solutions to inhibit their growth. However, the impedance at the operating frequency range is small enough that the capacitors do not have a significant impact on the accuracy of the method. 

The gain extrapolated from numerical results for the Table~\ref{table:JTWPA-RPM-params} JTWPA using (\ref{eq:gain}) is compared against the analytical result from \cite{2023_Yuan} in Fig. \ref{fig:gain-curve}. For this architecture, a fourth-order polynomial interpolating function was found to provide good agreement with the analytical results in the vicinity of the pump. At $6 \, \mathrm{GHz}$, the pole introduced by the resonant loads produces a large decrease in gain. Due to the use of interpolation in this region, this effect is not modeled by the numerical method.

\subsection{Josephson Junction Area Variations}
\label{subsec:JJ-variations}

To demonstrate a potential application of the numerical method, we investigated the effect of variations in Josephson junction area on JTWPA gain. The junction area $A$ is related to the Josephson inductance and capacitance by $L_{\mathrm{J}} \propto A$ and $C_{\mathrm{J}} \propto 1/A$ \cite{2004_Tinkham_Superconductivity}. Fig. \ref{fig:JTWPA-noRPM-areaVar}\subref{subfig:noRPM-areaGrad=0} demonstrates the result of independently varying the area of each Josephson junction in the Table \ref{table:JTWPA-params} JTWPA on a normal distribution with mean $A$ and standard deviation $\sigma_A$. For $\sigma_A = 0.03A$, the impact is relatively minor, and gain remains close to the analytical result. The differences are more pronounced for $\sigma_A = 0.06 A$ and $\sigma_A = 0.09 A$, especially at signal frequencies greater than the pump frequency, where substantial gain ripples occur. 

In addition to entirely random variations, significant gradients in Josephson junction area across a JTWPA have been observed during device manufacturing \cite{2019_Zorin_Flux-Driven}. To model these gradients, we varied the mean of the normal distribution as a linear function of position along the JTWPA. In Fig. \ref{fig:JTWPA-noRPM-areaVar}\subref{subfig:noRPM-areaGrad=5}, the gradient is set so that the mean area of the final Josephson junction in the JTWPA is $0.05 A$ higher than the first junction. In Fig. \ref{fig:JTWPA-noRPM-areaVar}\subref{subfig:noRPM-areaGrad=10} and Fig. \ref{fig:JTWPA-noRPM-areaVar}\subref{subfig:noRPM-areaGrad=15}, this procedure was repeated with total gradients of $0.10 A$ and $0.15 A$ respectively. These results indicate that increasing the gradient reduces gain across the bandwidth of the device.

\begin{figure}[t!]
\centering
  \includegraphics[width=0.9\linewidth]{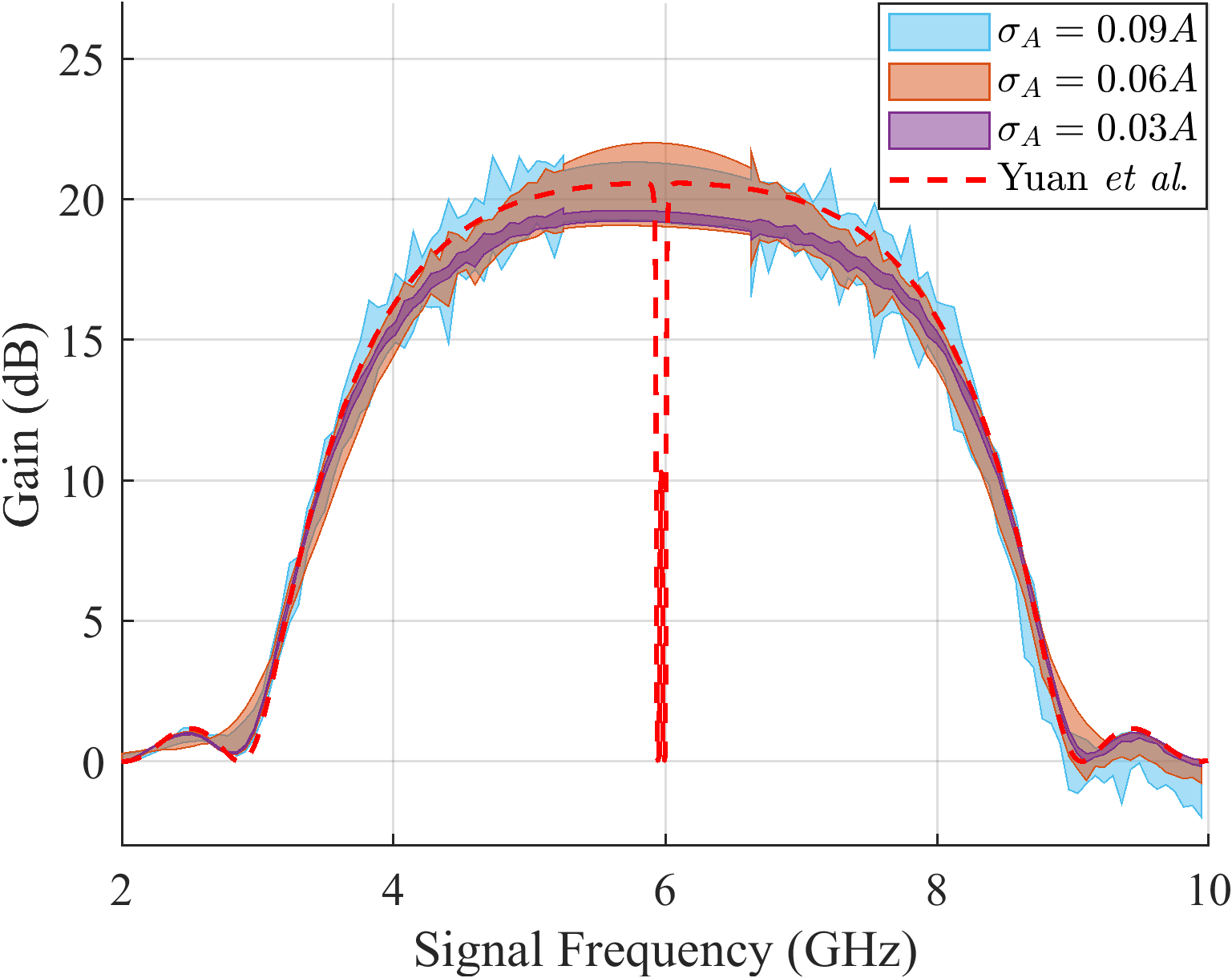}
  \caption{Gain of the Table \ref{table:JTWPA-RPM-params} JTWPA with variations in Josephson junction area. Each curve is shaded between the minimum and maximum outcomes of 30 simulations. The area of the Josephson junctions are selected from a normal distribution with standard deviation $\sigma_A$, and with its mean linearly varied by $0.05A$ across the JTWPA.}
  \label{fig:JTWPA-RPM-areaVar}
\end{figure}

The impact of junction area fluctuations was also tested for the JTWPA with RPM in Fig. \ref{fig:JTWPA-RPM-areaVar}, where a gradient of $0.05A$ was used. In this case, the presence of the resonant loads leads to additional gain ripples near the pump frequency.

\subsection{Resonant Load Variations}
\label{subsec:RPM-variations}

\begin{figure}[t]
\centering
  \includegraphics[width=0.9\linewidth]{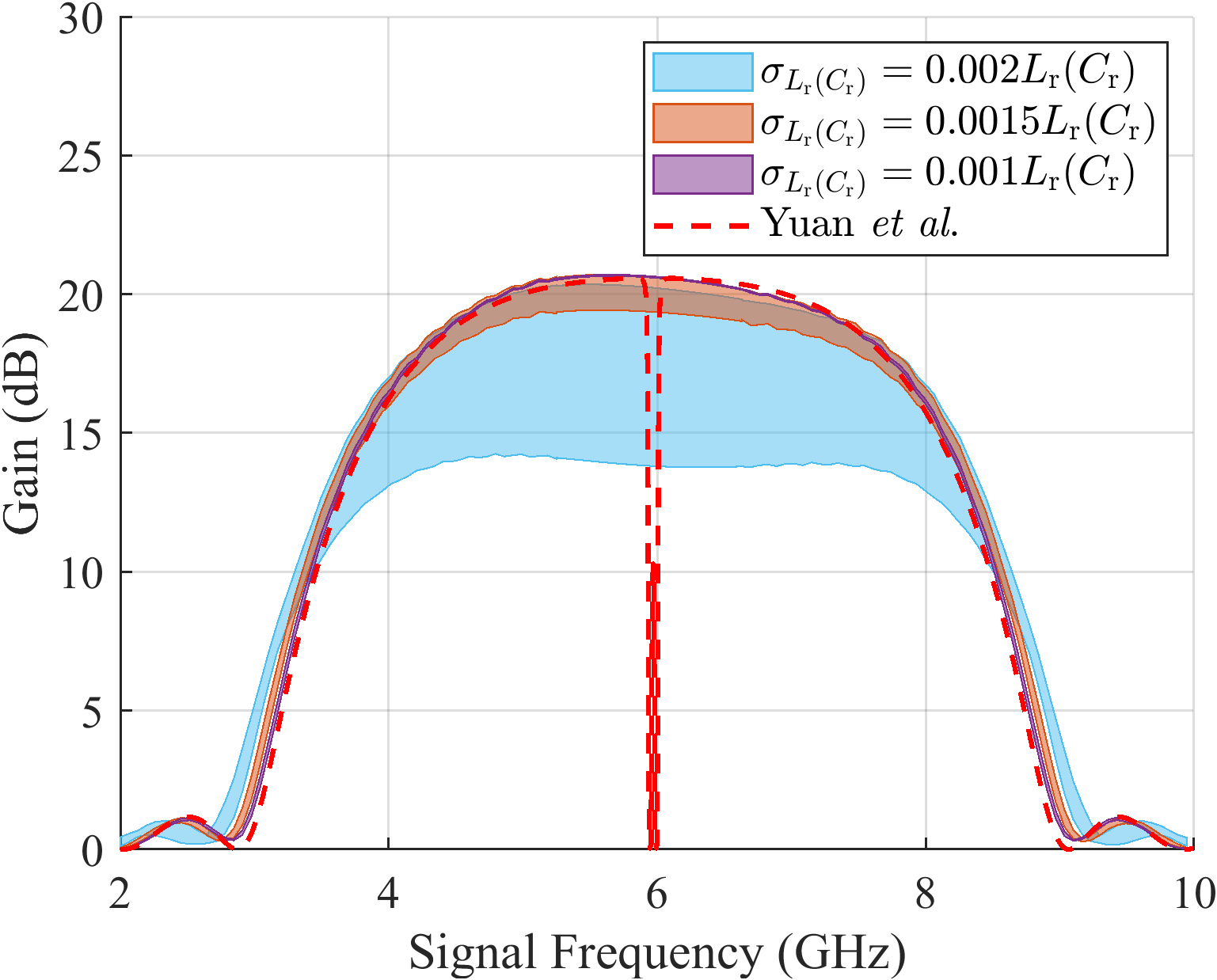}
  \caption{Gain of the Table \ref{table:JTWPA-RPM-params} JTWPA with variations in the resonant loads. Each curve is shaded between the minimum and maximum outcomes of 30 simulations. For each load, $L_\mathrm{r}$ and $C_\mathrm{r}$ are independently chosen from normal distributions with standard deviations $\sigma_{L_\mathrm{r}}$ and $\sigma_{C_\mathrm{r}}$, and mean values matching those in Table \ref{table:JTWPA-RPM-params}. }
  \label{fig:JTWPA-RPM-var}
\end{figure}

In addition to the Josephson junction area, the method was applied to investigate how small variations in the values of the resonant load components can affect JTWPA gain. While these components do not suffer from the same manufacturing inconsistencies as Josephson junctions, the delicate nature of the phase matching condition means small changes can severely impact performance. To demonstrate this, we varied $L_{\mathrm{r}}$ and $C_{\mathrm{r}}$ for each load on normal distributions with standard deviations $\sigma_{L_{\mathrm{r}}}$ and $\sigma_{C_{\mathrm{r}}}$. The mean value for each distribution matches the component value used in Table \ref{table:JTWPA-RPM-params}. The result of applying these variations are displayed in Fig. \ref{fig:JTWPA-RPM-var} for several values of $\sigma_{L_{\mathrm{r}}}$ and $\sigma_{C_{\mathrm{r}}}$. For $\sigma_{L_{\mathrm{r}} (C_{\mathrm{r}})} = 0.0010 L_{\mathrm{r}} (C_{\mathrm{r}})$ and $\sigma_{L_{\mathrm{r}} (C_{\mathrm{r}})} = 0.0015 L_{\mathrm{r}} (C_{\mathrm{r}})$, the effect of these variations is mostly negligible, indicating that changes in the pump phase due to the shifting resonance roughly cancel across the JTWPA. However, the impact at $\sigma_{L_{\mathrm{r}} (C_{\mathrm{r}})} = 0.0020 L_{\mathrm{r}} (C_{\mathrm{r}})$,  is clearly much more significant. At this value, the standard deviation is large enough that the pole may be shifted so that it overlaps with the pump bandwidth. As a result, the pump is partially absorbed by the resonant load, leading to the gain drop-off observed in the plot.

\section{Conclusion}
\label{sec:conclusion}

In this work, a numerical method for modeling JTPWAs was developed to overcome the limitations of analytical models. Rather than using a continuum approximation as typically done in analytical approaches, the Josephson junctions and transmission lines were treated as coupled subsystems. As a result, components can easily be individually varied to examine the impact of manufacturing tolerances on device performance. To linearize the time-marching equations for the dynamical variables, we employed a leap-frog time-marching procedure. Gain was extracted from simulations of JTWPA designs with and without RPM and compared to analytical results to validate the method. Further, the impact of variations in Josephson junctions and resonant load components was investigated using the method.

Future work will expand the method to support additional JTWPA architectures and explore other effects not captured by analytical models. 
Modern JTWPA designs integrate Josephson junction-based circuit elements such as superconducting quantum interference devices (SQUIDs) \cite{2020_Planat_Photonic_JTWPA} and superconducting nonlinear asymmetric inductive elements (SNAILs) \cite{2022_Ranadive} to leverage their additional degrees of freedom. These components can also be used to develop three-wave mixing JTWPAs which are inherently phase matched \cite{2015_Zorin_3-wave}, and can use a flux-driven topology to achieve high dynamic range \cite{2019_Zorin_Flux-Driven, 2024_Haider_Black-box}. However, the performance of these JTWPAs is heavily impacted by the generation of harmonics, an effect which is not captured by analytical models \cite{2020_Dixon_JTWPA_Complex_Behavior}. By incorporating components such as SQUIDs and SNAILs into our multiphysics method, this issue can be explored in greater detail.

In addition to supporting more JTWPA architectures, future work will extend the numerical method to a 3D full-wave description. While analytical device models are inherently limited to 1D, the need to extract circuit parameters of JTWPA geometries limits the accuracy of those methods and adds a significant pre-processing step. Meanwhile, a 3D numerical method would be capable of modeling the intended and non-ideal behavior of JTWPAs with unprecedented accuracy.




\bibliographystyle{IEEEtran}
\bibliography{citations}

\begin{IEEEbiography}[{\includegraphics[width=1in,height=1.25in,clip,keepaspectratio]{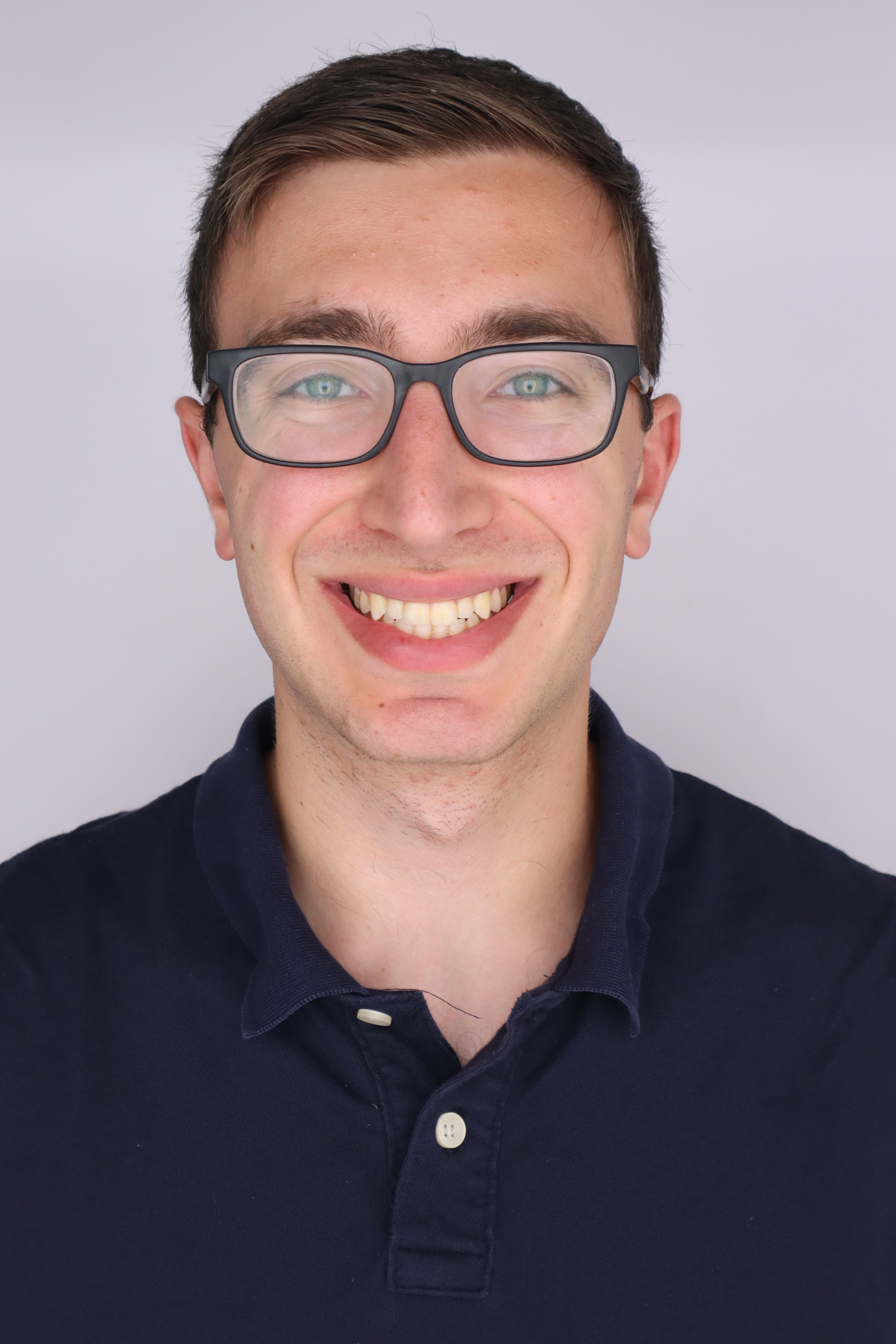}}]{Samuel T. Elkin}
	(S'22) received the B.S degree in electrical engineering from Purdue University, West Lafayette, IN, USA, in 2022. From 2022 to 2023, he worked at Indesign, LLC., Indianapolis, IN, USA, as an electrical engineer. He is currently pursuing a Ph.D. degree in electrical and computer engineering at Purdue University. His current research interests are in multiphysics modeling of superconducting circuit quantum devices.
\end{IEEEbiography}

\begin{IEEEbiography}[{\includegraphics[width=1in,height=1.25in,clip,keepaspectratio]{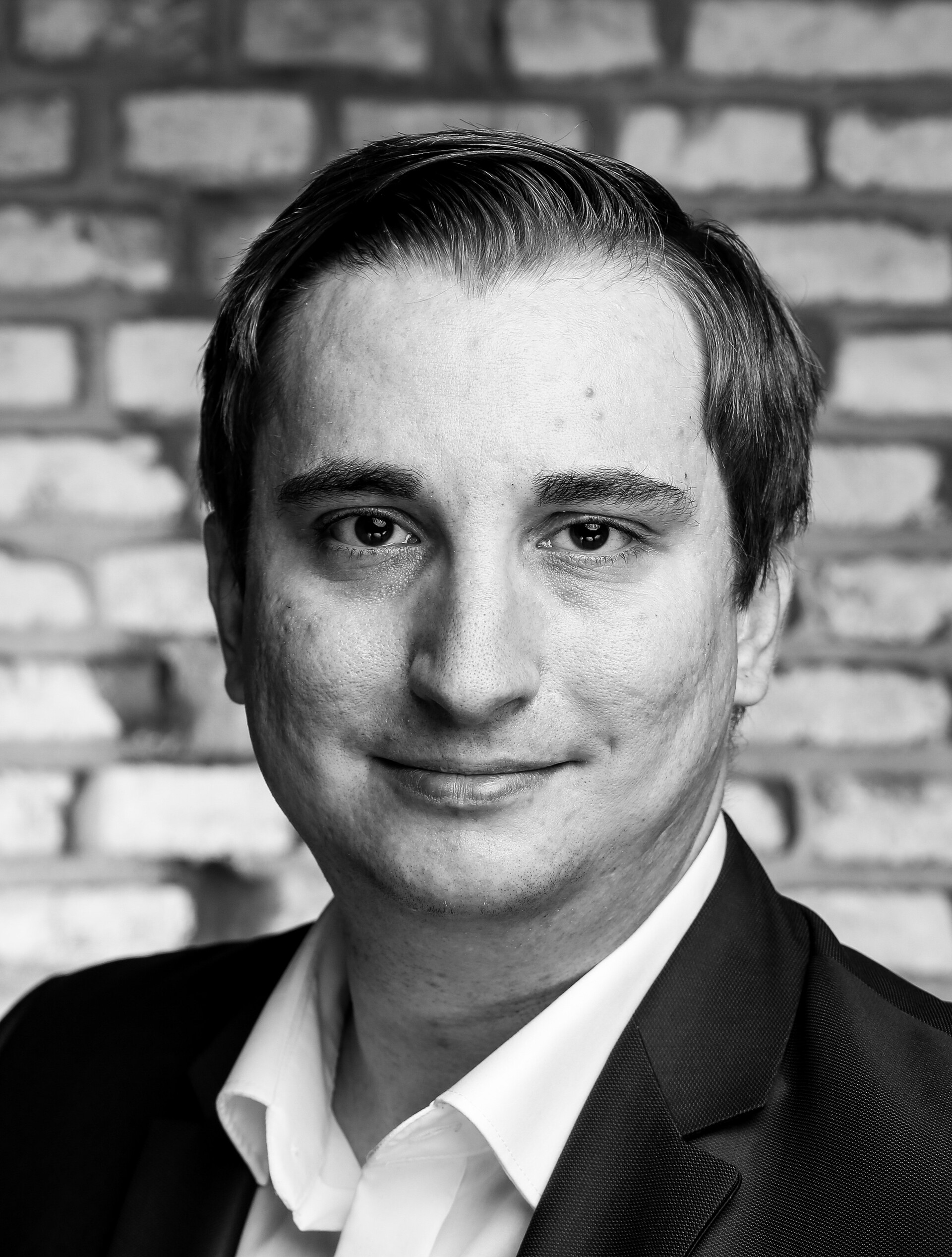}}]{Michael Haider} (Member, IEEE)
     received the B.Sc., M.Sc., and Ph.D. (summa cum laude) degrees in electrical and computer engineering from the Technical University of Munich (TUM), Garching, Germany, in 2014, 2016, and 2019, respectively. From 2016 to 2019, he was a Research Associate with the Institute of Nanoelectronics, TUM, where his research focused on quantum circuit theory, wireless power transfer, and stochastic electromagnetic fields. In 2019, he joined the Computational Photonics Group, TUM, as a Post-Doctoral Researcher, where he became a permanent staff Scientist in October 2023. He has coauthored over 100 scientific papers in refereed journals and conference proceedings. His current research interests include the modeling of charge carrier transport in quantum cascade lasers, the dynamical modeling of quantum devices within a generalized Maxwell-Bloch framework, and quantum circuit theoretical models applied to Josephson traveling-wave parametric amplifiers. Dr. Haider received the 2020 VDE ITG Dissertationspreis for his outstanding Ph.D. thesis on Investigations of Stochastic Electromagnetic Fields.
\end{IEEEbiography}

\begin{IEEEbiography}[{\includegraphics[width=1in,height=1.25in,clip,keepaspectratio]{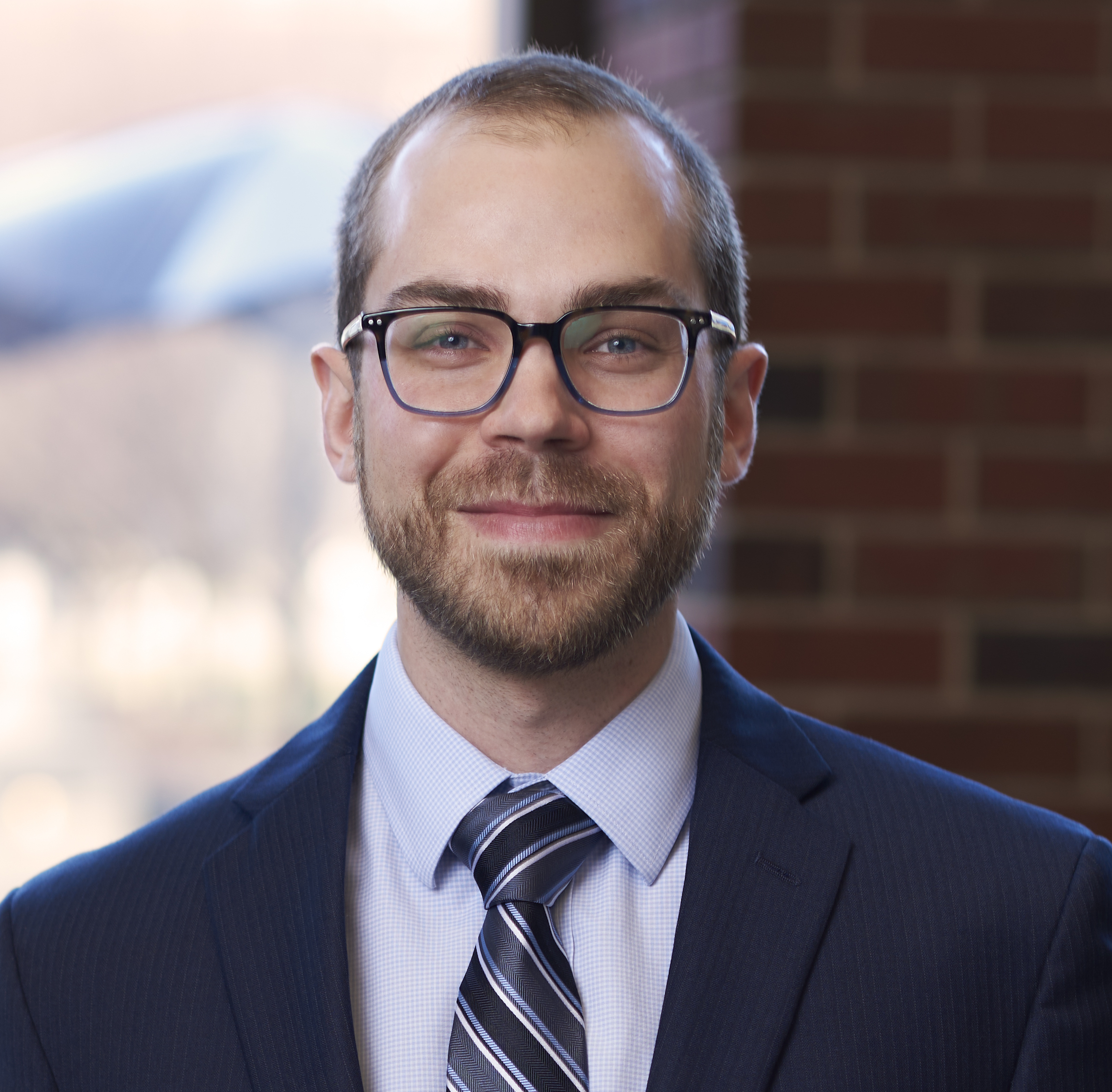}}]{Thomas E. Roth} (Member, IEEE)
	received the B.S. degrees in electrical engineering and computer engineering from the Missouri University of Science and Technology, Rolla, MO, USA, in 2015, and the M.S. and Ph.D. degrees in electrical and computer engineering from the University of Illinois at Urbana-Champaign, Champaign, IL, USA, in 2017 and 2020, respectively. He is currently an Assistant Professor with the Elmore Family School of Electrical and Computer Engineering, Purdue University, West Lafayette, IN, USA. Prior to joining Purdue, he was a Senior Member of the Technical Staff with Sandia National Laboratories, Albuquerque, NM, USA. He has received five best paper or young scientist awards from conferences, and the Ruth and Joel Spira Outstanding Teacher Award from Purdue University. His current research interests include multiscale and multiphysics computational electromagnetics techniques, particularly for rigorously analyzing quantum information processing devices.
\end{IEEEbiography}






\end{document}